\documentclass[aps,preprint,nofootinbib,preprintnumbers,eqsecnum,superscriptaddress]{revtex4}

\usepackage{amsmath} 
\usepackage{graphicx} 
\usepackage{amsthm}
\usepackage{amssymb} 
\usepackage{dsfont}
\usepackage{yfonts}
\usepackage{hyperref}
\usepackage{array,xcolor,graphicx}
\usepackage{booktabs,multirow}
\usepackage[utf8]{inputenc}
\usepackage{mathtools}
\usepackage{wrapfig}

\usepackage{subcaption}

\usepackage{slashed}
\usepackage{caption}
\usepackage{stackrel}
\usepackage{cases}
\usepackage{tcolorbox}
\usepackage{subcaption}

\usepackage{etoolbox}
\patchcmd{\section}
  {\centering}
  {\raggedright}
  {}
  {}
\patchcmd{\subsection}
  {\centering}
  {\raggedright}
  {}
  {}

\usepackage[normalem]{ulem}
\usepackage{mathtools}

\usepackage[all]{xy}
\usepackage{tikz}
\usetikzlibrary{arrows.meta}

\hypersetup{colorlinks=true,linkcolor=blue,citecolor=red,urlcolor=blue, linktocpage=true}

\newcommand{\be}{\begin{equation}}
\newcommand{\ee}{\end{equation}}
\newcommand{\bea}{\begin{eqnarray}}
\newcommand{\eea}{\end{eqnarray}}

\newcommand{\ka}{\kappa}

\newcommand{\bln}{\begin{align}}
\newcommand{\eln}{\end{align}}
\newcommand{\bst}{\begin{split}}
\newcommand{\est}{\end{split}}
\newcommand{\bi}{\begin{itemize}}
\newcommand{\ei}{\end{itemize}}
\newcommand{\ben}{\begin{enumerate}}
\newcommand{\een}{\end{enumerate}}

\def\le{\left}
\def\ri{\right}
\def\ha{{1\over 2}}

\def\vev#1{\langle#1\rangle}

\def\th{{\theta}}

\def\ep{{\epsilon}}

\newcommand{\p}{\partial}

\newcommand\sig{\sigma}

\newcommand\lam{\lambda}

\def\lam{{\lambda}}

\def\eeq{\end{equation}}

\newcommand\sH{{\ensuremath{{\mathcal H}}}}
\newcommand\sL{{\ensuremath{{\mathcal L}}}}

\newcommand\sO{{\ensuremath{{\mathcal O}}}}

\newcommand\tsig{\tilde{\sigma}}

\def\th{{\theta}}

\DeclareMathOperator*{\argmin}{arg\,min}
\usepackage{algorithm}
\usepackage{algpseudocode}

\numberwithin{equation}{section}

\begin{document}

\title{Machine learning and optimization-based approaches to duality in statistical physics}
\author{Andrea E. V. Ferrari}
\email{andrea.e.v.ferrari@gmail.com}
\affiliation{Deutsches Elektronen-Synchrotron DESY, Notkestraße 85, 22607 Hamburg, Germany}
\affiliation{School of Mathematics, The University of Edinburgh, Mayfield Road, Edinburgh, U.K}
\author{Prateek Gupta}
\affiliation{Max Planck Institute for Human Development, Berlin, Germay}
\author{Nabil Iqbal} 
\email{nabil.iqbal@durham.ac.uk}
\affiliation{Centre for Particle Theory, Department of Mathematical Sciences, Durham University,
		South Road, Durham DH1 3LE, UK\\} 
  \affiliation{Amsterdam Machine Learning Lab,
University of Amsterdam, Science Park 900,
1098 XH Amsterdam, NL}
\begin{abstract}
The notion of {\it duality} -- that a given physical system can have two different mathematical descriptions -- is a key idea in modern theoretical physics. Establishing a duality in lattice statistical mechanics models requires the construction of a dual Hamiltonian and a map from the original to the dual observables. By using simple neural networks to parameterize these maps and introducing a loss function that penalises the difference between correlation functions in original and dual models, we formulate the process of duality discovery as an optimization problem. We numerically solve this problem and show that our framework can rediscover the celebrated Kramers-Wannier duality for the 2d Ising model, reconstructing the known mapping of temperatures. We also discuss an alternative approach which uses known features of the mapping of topological lines to reduce the problem to optimizing the couplings in a dual Hamiltonian, and explore next-to-nearest neighbour deformations of the 2d Ising duality. We discuss future directions and prospects for discovering new dualities within this framework. 
\end{abstract} 
\maketitle

\tableofcontents

\section{Introduction and background}
A key concept in physics is {\it duality}, i.e. the idea that the same physical system can have two different mathematical descriptions. Duality sits at the heart of modern theoretical physics. Some influential examples selected from different areas of physics include \cite{Savit:1979ny,PhysRev.60.252,Kramers:1941zz,Peskin:1977kp,dasgupta1981phase,Coleman:1974bu,Wegner1971,Montonen:1977sn,Seiberg:1994rs,Maldacena:1997re}.

In this work we seek to formalize the task of finding a duality in the controlled setting of lattice statistical physics, in a manner that allows modern machine learning as well as other numerical techniques to be used to systematically search for dualities. Our ultimate hope is that completely novel dualities may be found using such approaches. We will not realize these hopes in this work, however. Rather, we will elucidate interesting technical challenges that such a formulation poses, and build two algorithms that are sufficiently flexible and robust to ``rediscover'' the celebrated Kramers-Wannier duality of the classical 2d Ising model~\cite{PhysRev.60.252,Kramers:1941zz} in a fully automated way.

The first algorithm, which we present in Section~\ref{sec:gen-neur}, is very general and has only elementary definitions as input. It could therefore apply \emph{mutatis mutandis} to a variety of physical systems, perhaps even beyond the controlled setting of duality in lattice models. The second algorithm, which we present in Section~\ref{sec:top-lines}, relies instead on more refined but still rather elementary physical arguments (in particular the presence of topological lines associated with discrete symmetries \cite{Gaiotto:2014kfa}), and could efficiently be exploited to perform numerical explorations of dualities of Kramers-Wannier type. We will demonstrate this by studying deformations of the classical 2d Kramers-Wannier duality by the introduction of nearest-neighbour couplings, leaving further applications to future work.

A portion of this work (the contents of Section \ref{MLapproach}, where the first algorithm is discussed) will appear as a workshop paper at the Machine Learning for the Physical Sciences Workshop at NeurIPS2024.

\subsection{Background on duality}  \label{sec:dualitybg} 
We begin by formulating the notion of duality as we will use it in this work, and by formalizing the task of finding dualities.

Consider a physical system described by some degrees of freedom $\phi_i$, where generally $i \in L$ labels points in some (possibly discrete) space $L$, e.g. a regular lattice of spatial points. We are interested in classical statistical physics, where typically the dynamics of the system is described by a Hamiltonian (or equivalently, a Euclidean action) functional $H[\phi]$ which associates a probability to each field configuration $p[\phi] = Z^{-1} \exp(-H[\phi])$, where we have defined the partition function $Z \equiv \int [d\phi] \exp(-H[\phi])$. We are often interested in observables such as the correlation functions of the field $\phi_i$, i.e.
\be\label{eq:corr-def}
\langle \phi_{i_1} \phi_{i_2} \cdots \phi_{i_n} \rangle_{H} = \frac{1}{Z} \int [d\phi] \phi_{i_1} \phi_{i_2} \cdots \phi_{i_n} \exp(-H[\phi])
\ee
Here the subscript $H$ denotes that we take the average with respect to the Hamiltonian $H[\phi]$. 

Imagine now that we have a different set of degrees of freedom which we call $\tilde{\phi}_{\tilde{i}}$, where $\tilde{i}\in \tilde{L}$ labels points of what could in principle be an entirely different space $\tilde{L}$\footnote{In the case of gauge/gravity duality -- outside the scope of this work -- it even has a different dimensionality.}, and with dynamics described by their own Hamiltonian $\tilde{H}[\tilde{\phi}]$. In analogy with~\eqref{eq:corr-def}, $\langle \tilde{\phi}_{\tilde{i}_1} \cdots \tilde{\phi}_{\tilde{i}_n}\rangle_{\tilde{H}}$ denotes a correlation function of the new variables weighted by their own action $\tilde{H}[\tilde{\phi}]$.

A {\it duality} as understood here arises when there is a way to obtain the observables of the system described by $\phi_i$ from those described by $\tilde{\phi}_{\tilde{i}}$. For example, for each function $F(\{\phi_i\})$ of the old variables there must be some function $G(\{\tilde{\phi}_{\tilde{i}}\})$ of the new variables such that
\be\label{eq:basic-dual}
\langle F(\{\phi_i\}) \rangle_{H} = \langle G(\{\tilde{\phi}_{\tilde{i}}\}) \rangle_{\tilde{H}} \ . 
\ee

% \langle \phi_{i_1} \phi_{i_2} \cdots \phi_{i_n} \rangle_{H} = \langle G(\tilde{\phi}_{\tilde{i}_1},\tilde{\phi}_{\tilde{i}_2}\cdots \tilde{\phi}_{\tilde{i}_m}) \rangle_{\tilde{H}} \ . 
% \ee
Note that the function $G$ may be quite complicated, and it may be possible for one of the sides to be local in space (i.e. to involve only a small number of widely separated $\phi_i$) while the other is non-local (e.g. it may involve extended objects which involve many $\tilde{\phi}_{\tilde{i}}$). If a mapping such as the above can be obtained for {\it all} physical observables, then we say that the two systems are {\it dual}, i.e. they provide two different mathematical representations of the same physical theory.

 In practice, it may be sufficient to restrict our attention to a limited set of elementary functions $F_a(\{\phi_i\})$ with the index $a$ taking values in some controlled set, and not any arbitrary function $F$. (In a gauge theory, these may be gauge-invariant functions that are polynomials in the degrees of freedom, or taken from another reasonable class.) For agreement of these elementary functions to constitute a full duality, it must be possible to approximate any reasonable $F$ by taking products and sums of the elementary functions. This will impose some consistency conditions for the duality mapping. For instance, products of these elementary functions should be sent to products of their duals,
\be\label{eq:mapping-intro}
\langle F_1(\{\phi_i\}) \cdots F_n(\{\phi_i\}) \rangle_{H} = \langle G_1(\{\tilde{\phi}_{\tilde{i}}\})  \dots G_n(\{\tilde{\phi}_{\tilde{i}}\}) \rangle_{\tilde{H}} \ . 
\ee
In statistical jargon, this implies in particular that all moments of the elementary functions and their duals agree. This will be the kind of duality considered in this paper. Concrete examples will be given below. 

We can then formalize the task of finding such a duality. In its most general form, the task reads as follows. Given an original system determined by some space $L$, degrees of freedom $\phi_i$ and Hamiltonian as above, we want to construct the following objects:
\begin{itemize}
\item A choice of space $\tilde{L}$,
\item A choice of dual variables $\tilde{\phi}_{\tilde{i}}$ where $\tilde{i}\in \tilde{L}$,
\item An Hamiltonian functional $\tilde{H}[\tilde{\beta}, \tilde{\phi}_{\tilde{i}}]$ which associates probabilities to field configurations, parametrized by a set of couplings $\tilde{\beta}$, and 
\item For each selected elementary function $F_a(\{\phi_i\})$ of the original variables, a mapping $G_a(\{\tilde{\phi}_{\tilde{i}}\})$ so that \eqref{eq:mapping-intro} holds. 
\end{itemize}
A genuine duality is found whenever any of the objects above differs from the original formulation.  

In this paper we will focus on dualities of lattice models, where $\phi_i = \sigma_i$ are some ``spin" variables labeled by lattice sites $i\in L$. A prototypical example, which we will focus on in this work, is Kramers-Wannier duality for the 2d Ising model \cite{PhysRev.60.252,Kramers:1941zz}. The model consists of spins $\sig_i = \pm 1$ where $L\cong \mathbb{Z}^2$ is in this case a square lattice, with Hamiltonian $H$ and partition function $Z$:
\be
H[\beta,\sig_i] = -\beta \sum_{\vev{ij}} \sig_i \sig_j \ \qquad Z = \sum_{\sig_i} \exp(-H) \label{isingH} 
\ee
The probability of obtaining a given spin configuration is given by the Boltzmann weight $p[\sig] = Z^{-1} \exp(-H[\sig])$. There is a $\mathbb{Z}_2$ global symmetry which acts on the spins as $\sig_i \to -\sig_i$. 
\begin{figure}
    \centering
    %\vspace{-0.75cm} 
    \includegraphics[width=0.3\linewidth]{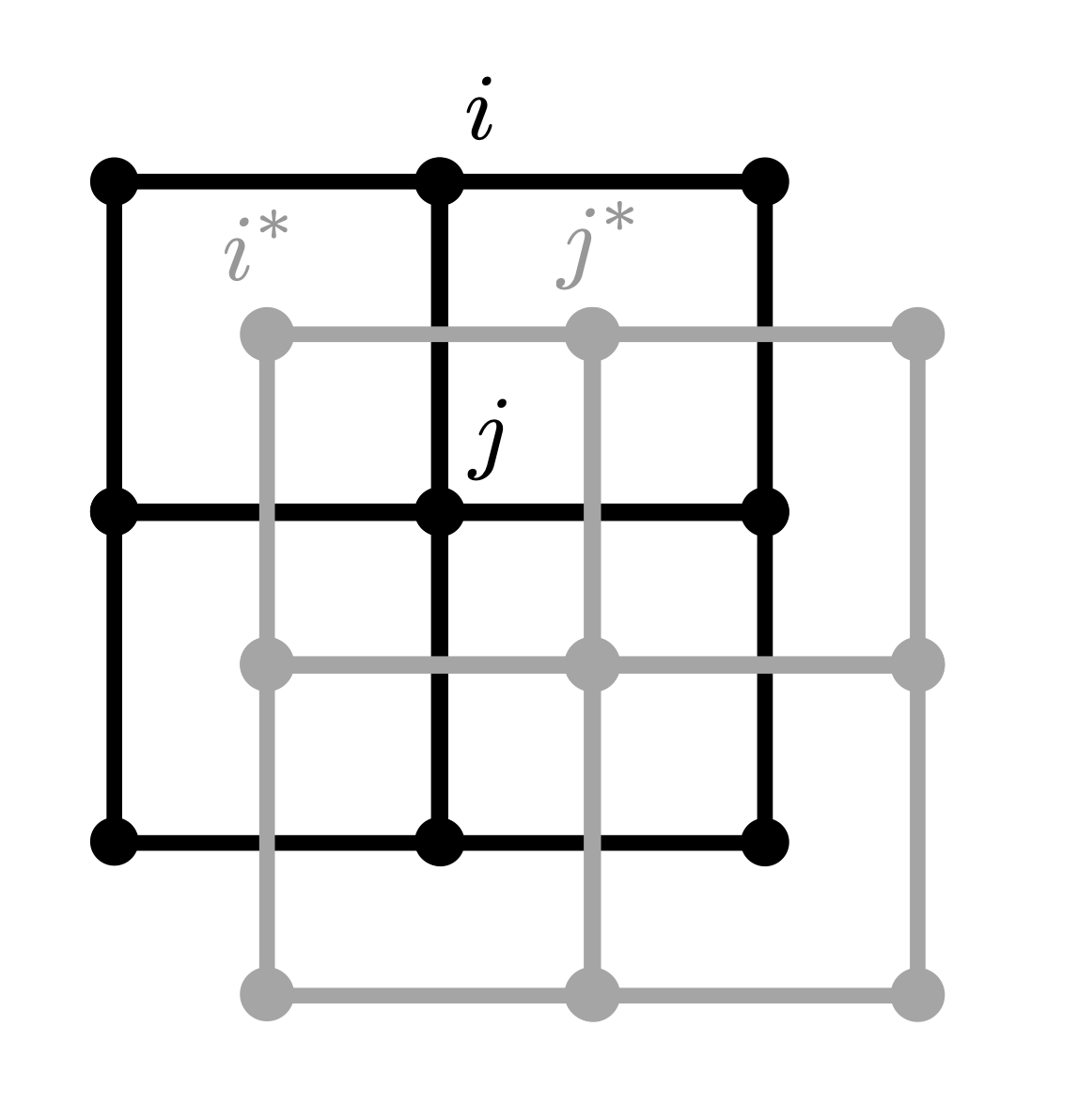}
    \caption{The two-point product of spins in the original frame $\sig_i \sig_j$ is related to the product of spins $\tilde{\sig}_{i^*} \tilde{\sig}_{j^*}$ in the dual frame, where $i^*,j^*$ are related to $ij$ as shown.}
    %\NI{explain rot}
    \label{fig:duallink}
\end{figure}
It is now a remarkable fact (and essentially a consequence of gauging the $\mathbb{Z}_2$ symmetry\footnote{This can be interpreted as a Fourier transform, related to electro-magnetic duality for finite gauge groups in three dimensions~\cite{Freed:2018cec}.}) that the system described by \eqref{isingH} is precisely equivalent to a different system with spins $\tilde{\sig}_{\tilde{i}} = \pm 1$ living on the {\it dual} lattice, $\tilde{L}\cong L^\vee$ where again $\tilde{L} \cong \mathbb{Z}^2$:
\be
\tilde{H}[\tilde{\beta},\tilde{\sig}_{\tilde{i}}] = -\tilde{\beta} \sum_{\vev{\tilde{i}\tilde{j}}} \tilde{\sig}_{\tilde{i}} \tilde{\sig}_{\tilde{j}} \qquad \tilde{Z} = \sum_{\tilde{\sig}_{\tilde{i}}} \exp(-\tilde{H})~.
\label{dualH}
\ee
Here the inverse temperature $\tilde{\beta}$ is related to $\beta$ by
\be
\sinh(2\beta) \sinh(2\tilde{\beta}) = 1 \ . 
\ee

We present a self-contained review of this duality in Appendix \ref{app:KW}. All observables constructed from functions $F(\{\phi_i\})$ of the $\sig_i$ can be mapped to observables constructed from functions $G(\{\tilde{\phi}_{\tilde{i}}\})$ of the $\tilde{\sig}_{\tilde{i}}$. The mapping between these $F$'s and $G$'s is simplest when considering the $\mathbb{Z}_2$ even sector, where it is local. Take a {\it link product} 
\be
F_{ij} = \sigma_i\sigma_j\, \label{linkprod} 
\ee
where $i,j\in L$ labels two neighbouring sites in the original lattice (by translation-invariance of the original model, we can pick any such pair of sites for the following discussion). Then we have
\be
\vev{\sig_i \sig_j\cdots}_{\beta} = \big{\langle}e^{-2\tilde{\beta} \tilde{\sig}_{i^*} \tilde{\sig}_{j^*}}\cdots\big{\rangle}_{\tilde{\beta}} \label{mapping} 
\ee
where the notation $i^*$ refers to sites on the dual lattice $L^\vee$ such that the link connecting sites $i^*$ and $j^*$ intersects the the link connecting $i$ and $j$, as shown in Figure \ref{fig:duallink}. The $\cdots$ indicate that this is an operator equation which holds for arbitrary insertions of operators and thus can be used to construct any expectation value of an even number of the $\sig_i$. (An exception when we consider precisely the same two-spin operator {\it twice}, i.e. $(\sig_i \sig_j)^2 = 1$, when a careful derivation of the duality shows that the right-hand side must be modified and is also identically $1$.) Thus, in this case, we take our elementary functions to be of the form $F_{ij} = \sigma_i \sigma_j$ for all neighbouring spins, and our task is to find the dual lattice, the dual Hamiltonian and one function 
\be\label{mapping-is}
G_{ij}( \{\tilde{\sigma}_{\tilde{i}}\} ) = e^{-2\tilde{\beta} \tilde{\sig}_{i^*} \tilde{\sig}_{j^*} }~,
\ee
which is the dual to out elementary function $F_{ij}=\sigma_i\sigma_j$. This example is a special case of duality where the functional form of the dual Hamiltonian is identical to that of the original, but with a different value of the coupling $\tilde{\beta}$ and living on a different (dual) lattice.

\subsection{Relation to previous work} 

The problem of starting from data of the form
\be
\langle \phi_{i_1} \phi_{i_2} \cdots \phi_{i_n} \rangle_{H}
\ee
and learning the parameters $\beta$ entering a known Hamiltonian $H[\beta, \phi_i] $ is precisely that of training a Boltzmann machine. The problem has a very long and rich history (see e.g. \cite{bengio2017deep} for an overview)\footnote{As of this writing this history now includes the award of the 2024 Nobel Prize in Physics \cite{Nobelprize.org2024}.}.
%For the specific case of the Ising model architecture various independent, efficient solutions based on self-consistency conditions of the model have been developed in the context of the ``inverse Ising Problem'' (see 

Our case differs from this classical situation in that even though we can make a simple ansatz for a local Hamiltonian,
\begin{enumerate}
    \item we do not want to assume complete knowledge of the space $L$,
    \item we do not want to assume complete knowledge of the observables whose correlation function are being observed, i.e. we have to construct an appropriate mapping function $G_{a}(\{\tilde{\phi}_{\tilde{i}}\})$ as in \eqref{eq:mapping-intro}. 
\end{enumerate}
Simultaneously learning the parameters in the Hamiltonian, some structure of the space, and a function encoding the types of correlation functions being observed leads of course to a much harder problem. We demonstrate how to partially solve this problem at least in a controlled set-up.

Other work on dualities in physics involving machine learning includes \cite{Betzler:2020rfg,Bao:2020nbi,Capuozzo:2024vdw}, but none is aimed at recovering the full dual descriptions as formulated here.

\section{General neural network approach} \label{MLapproach} 
\label{sec:gen-neur}

\subsection{Methodology}

We now explain how, starting from the Hamiltonian $H[\beta,\sigma]$ of some statistical model on a lattice $L$, we can learn candidates $\tilde{H}[\tilde{\beta},\tilde{\sigma}]$ for dual models as well as a dictionary between original and dual observables. This dictionary will turn out to include a mapping from the original to the dual lattice, where our approach will allow for an arbitrary local map of observables on links that will ultimately (at the end of the network training) be interpreted as defining the dual lattice from the original one. 

%This includes learning the fact that the dual model is defined on a different lattice, such as the dual lattice $L^\vee$; for simplicity, we take $L$ to be a square lattice, whose dual is also a square lattice (but related to it in a nontrivial way, see Figure~\ref{fig:duallink}). For simplicity, we will introduce the freedom of working on either the original or dual lattice 

\begin{wrapfigure}[16]{r}{0.4\textwidth}
    \centering
    \vspace{-0.5cm} 
    \includegraphics[width=0.95\linewidth]{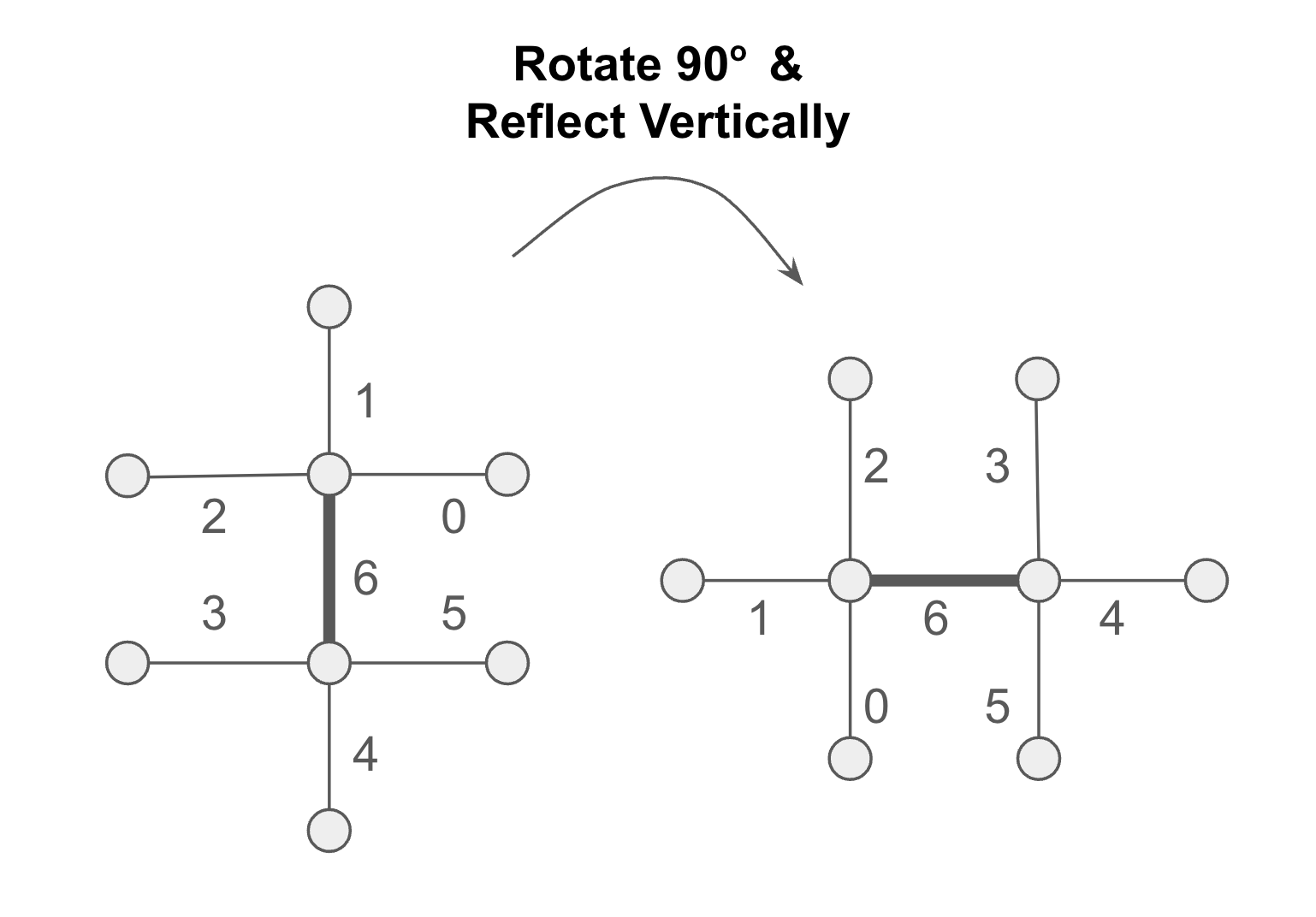}
    \caption{We parametrize $G$ as a neural network that takes neighbouring links of a given link (in this case \# 6) as its input. The assignment on horizontal links is related to that on vertical ones by a rotation and reflection.}
    %\NI{explain rot}
    \label{fig:linkMap}
\end{wrapfigure}

%\subsection{Framework and loss function}  
\paragraph{Framework and loss function:} \label{sec:framework}
%\begin{figure*}[htp!]
We assume that $\tilde{H}$ can be written in terms of local couplings of spins:
\be
\tilde{H}[\tilde{\beta},\tilde{\sig}_{i}] = -\tilde{\beta} \sum_{\vev{ij}} \tilde{\sig}_{i} \tilde{\sig}_{j} - \tilde{\beta}_4 \sum_{\vev{ijkl}} \tilde{\sig}_{i} \tilde{\sig}_{j} \tilde{\sig}_{k} \tilde{\sig}_{l} - \cdots\label{dualH}
\ee
where the couplings $\tilde{\beta}_b$ are  parameters to be learned. We would like to find dual representations of the link products \label{linkprod} $F_{ij} = \sig_i \sig_j$  we described for the Ising model. We assume that the link product in the original model is mapped to \emph{some} functions of {\it nearby} link products in the dual model, more precisely
\be
F_{ij}\mapsto  G_{ij}( \{\tilde{\sigma}_{k} \tilde{\sigma}_{l} \}  ) 
\ee
where $\{\tilde{\sigma}_{k} \tilde{\sigma}_{l} \}$ is a set of link products such as the one shown in Figure \ref{fig:linkMap}. We will use the simplified notation $G(\tilde{\sigma})$ below.

$G$ is designed to be sufficiently flexible to recover models on lattices related in various ways to the original one. Note that a choice must be made about how to relate the assignment of link products neighbouring a horizontal link to the assignment of link products neighbouring a vertical link, as multiple choices are consistent with rotational invariance. In Figure \ref{fig:linkMap} we display the choice used, which relates them by a rotation composed with a reflection. As we will see later, this choice is important for recovering the geometry of the dual lattice.
%\AF{Andrea / Nabil: (R2) Describe the rotation and reflection more clearly. (R5) In particular, the choice of features on l.77 is motivated with an appeal to Figure 2, it wasn't clear how the same geometry could be attributed to both the primal and dual lattices.  The second term in the definition of $\ell^a$ is for the same feature as in the first term, but presumably the optimisation step seeks to find the closest dual set of edges that approximates the first?  (Apart from recovering the original model itself.} \NI{Added some more words here}. 

\begin{wrapfigure}[8]{r}{0.45\textwidth}
    \centering
\includegraphics[width=0.7\linewidth]{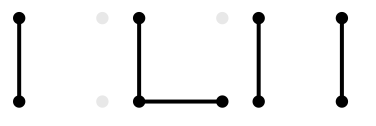}
    \caption{Examples of three features showing link products considered.}
    \label{fig:features}
\end{wrapfigure}

We now construct a loss function $\sL$ that is minimized when all correlation functions of $F_{ij}$ and $G_{ij}$ agree on the two sides of the duality, as in~\eqref{eq:mapping-intro}. This is similar to the matching of moments of two distributions, which is a standard problem, and for which one can construct general kernels that are minimized only when all of the moments of two distributions agree (see e.g. \cite{li2015generative}). Unfortunately, in the present case we cannot use such kernels because of one conceptual and one technical problem:
\begin{enumerate}
\item{
As discussed below \eqref{mapping} certain moments should not be matched, i.e. those which involve repeated moments of the {\it same} spin and thus contain a factor which is identically one: $(\sig_i)^2 = 1$. }
\item{
No notion of locality is embedded in standard moment matching (in the present case, correlation functions of faraway spins carry little information due to the exponential decay of correlations), and attempting to match them is a waste of computation. }
% Unfortunately, we found that these kernels did not work well for our purposes. We believe this arises from the fact that spatial locality is not manifest in those kernels: correlation functions of faraway spins carry no information due to the exponential decay of spatial correlations, and attempting to match them across the duality is a waste of computation. 
\end{enumerate}
Instead we explicitly match {\it features} -- i.e. moments of a small number of nearby link products, as shown in Figure~\ref{fig:features}  -- which we then spatially average over the lattice. Denoting these features as $\psi^{a}$ with $a$ running over features, we then construct the loss
\be
\sL(G,\tilde{H}) = \sum_{a} \ell^a \ell^a 
\qquad \ell^a = \vev{\psi^{a}[\sig_i]}_H - \vev{\psi^{a}[G(\tilde{\sig}_{i})]}_{\tilde{H}}
\label{lossdef} \ee
$\ell^a$ can be thought of as a vector in feature space indicating how far apart the two theories are.

It is clear that this loss can be minimized in two scenarios:  (a) $\tilde{H} = H$, $G(\tilde{\sig}_{i}\tilde{\sig}_{j}) = \tilde{\sig}_{i}\tilde{\sig}_{j}$ i.e., the original model is rediscovered, or (b) $\tilde{H} \neq H$, $G(\tilde{\sig}_{i}\tilde{\sig}_{j}) \neq \tilde{\sig}_{i}\tilde{\sig}_{j}$ representing a nontrivial dual model where (selected) moments nevertheless perfectly match those of the original model.\\

\paragraph{Optimization:}
We now need to solve the following optimization problem:
\be
G^*, \tilde{H}^* = \argmin_{G, \tilde H} \mathcal{L}(G, \tilde{H})
\label{optim-framework}
\ee
$G$ is represented by a neural network with parameters $\th$, $G = G_{\th}$. 

Algorithm~\ref{algo:opt} outlines the procedure for optimization.
Given a trial set of parameters $\th$ and couplings for the dual Hamiltonian $\tilde{\beta}_a$, we simultaneously perform Markov Chain Monte Carlo (MCMC) sampling from the original and dual Hamiltonians using a standard Metropolis algorithm to obtain spin configurations $\sig_i$ and $\tilde{\sig}_{\tilde{i}}$ drawn from the appropriate distributions respectively. We can then evaluate the expectation values in \eqref{lossdef}, and compute the loss $\sL$.

To minimize it we also need to compute gradients $\p_{\th} \sL$ and $\p_{\tilde{\beta}_a}\sL$. For $\th$ this can be done straightforwardly using conventional automatic differentiation techniques. For the $\tilde{\beta}_a$ we cannot backpropagate through a stochastic sampler, but explicit differentiation shows that we can relate the gradients to expectation values as follows. For concreteness we demonstrate the argument with only a single nonzero coupling $\tilde{\beta}$ in \eqref{dualH}, but the generalization to other couplings is immediate. For any function of spins $\sO[\tilde{\sig}]$ we have
\be
\vev{\sO}_{\tilde{H}} \equiv \frac{1}{Z(\tilde{\beta})} \sum_{\{\tilde{\sig}_{i}\}} \sO[\tilde{\sig}] e^{\le(\sum_{\vev{ij}}\tilde{\beta} \tilde{\sig}_{i} \tilde{\sig}_{j}\ri)} \qquad Z(\tilde{\beta}) \equiv \sum_{\{\tsig_i\}} e^{\le(\sum_{\vev{ij}}\tilde{\beta} \tilde{\sig}_{i} \tilde{\sig}_{j}\ri)} \label{Odef} 
\ee
where the sum over $\{\sig_i\}$ runs over all spin configurations. Now we have
\be
\p_{\tilde{\beta}} \sL = - 2\sum_{a} \ell^{a} \p_{\tilde{\beta}} \vev{\psi^a[G(\tilde{\sig}_{i})]}_{\tilde{H}},
\ee
where we have used the definition of $\ell^{a}$ in \eqref{lossdef}. 
From \eqref{Odef} the gradient of any observable with respect to $\tilde{\beta}$ is
\be
\p_{\tilde{\beta}}\vev{\sO}_{\tilde{H}} = -\vev{\sO}_{\tilde{H}} \sum_{\vev{ij}} \vev{\tsig_{i} \tsig_{j}}_{\tilde{H}} + \sum_{\vev{ij}} \langle{ \tsig_{i} \tsig_{j} \sO\rangle}_{\tilde{H}} \label{gradL} 
\ee
where the first term comes from differentiating $Z(\tilde{\beta})$ and the second from differentiating inside the Boltzmann measure weighting each configuration in \eqref{Odef}. Using this expression to evaluate \eqref{gradL} for $\sO = \psi^{a}[G(\tsig_i)]$ we find:
\be
\p_{\tilde{\beta}} \sL = 2\bigg\langle\sum_a \ell^a \le(\sum_{\vev{ij}}  \vev{ \tilde{\sig}_{i}\tilde{\sig}_{j}}_{\tilde{H}} - \sum_{\vev{ij}} \tilde{\sig}_{\tilde{i}} \tilde{\sig}_{j} \ri)\phi^a[G_{\th}(\tilde{\sig})]\bigg\rangle_{\tilde{H}} \ .  \label{gradexp} 
\ee
We can now evaluate the expectation value by MCMC sampling from the dual Hamiltonian. We note that this evaluation is computationally expensive, as each gradient step requires us to equilibrate an MCMC chain. For training conventional Boltzmann machines, one can use more efficient approaches such as contrastive divergence \cite{carreira2005contrastive}. Due to the presence of the mapping $G$, we are not aware of a similarly efficient algorithm in our case, and indeed, all likelihood-based approaches seem conceptually difficult.

% \begin{wrapfigure}{r}{0.2\textwidth}
\begin{algorithm}[H]
\caption{Machine learning for finding statistical mechanics duality}
\label{algo:opt}
\begin{algorithmic}[1]
    \State \textbf{Inputs:} $\beta$, $\eta$ (learning rate), $N$ (number of samples)
    \State \textbf{Initialize:} $\tilde{\beta}_0 \in \mathbb{R}$, $\theta \in \mathbb{R}^d$
    \For{each epoch $t = 1, 2, \dots, T$}
        \State Draw $N$ samples $\{\sig_i\}_{i=1}^{N} \sim p(\sig | \beta)$
        \State Draw $N$ samples $\{\tilde{\sig}_{i}\}_{\tilde{i}=1}^{N} \sim p(\tilde{\sig} | \tilde{\beta})$ where $\tilde{\beta} \neq \beta$
        \State Compute the loss $\mathcal{L} = \frac{1}{N} \sum_{i=1}^{N} \mathcal{L}(\sig_i, G_{\theta}(\tilde{\sig}_{i}))$
        \State Compute the gradients $\p_{\tilde{\beta}} \mathcal{L}$ and $\p_{\theta} \mathcal{L}$
        \State Update the parameters:
        \begin{align*}
            \tilde{\beta}_{t+1} &\gets \tilde{\beta}_t - \eta \p_{\tilde{\beta}} \mathcal{L} \\
            \theta_{t+1} &\gets \theta_t - \eta \p_{\theta} \mathcal{L}
        \end{align*}
        \If{$\mathcal{L}$ has not improved for the last $X$ epochs}
            \State \textbf{Stop the optimization}
        \EndIf
    \EndFor
\end{algorithmic}
\end{algorithm}

\subsection{Experiments}
%In this section, we conduct experiments to establish KW duality using our framework.
%We define an appropriate objective function for matching all sets of moments. 
%Further, we define the neural network architecture. 
%Finally, we empirically illustrate the construction of duality using the plots showing the progression of learning methods.

%\begin{figure}{l}{0.7\textwidth}
\begin{figure*}[htp!]
    \centering
    \includegraphics[width=\linewidth]{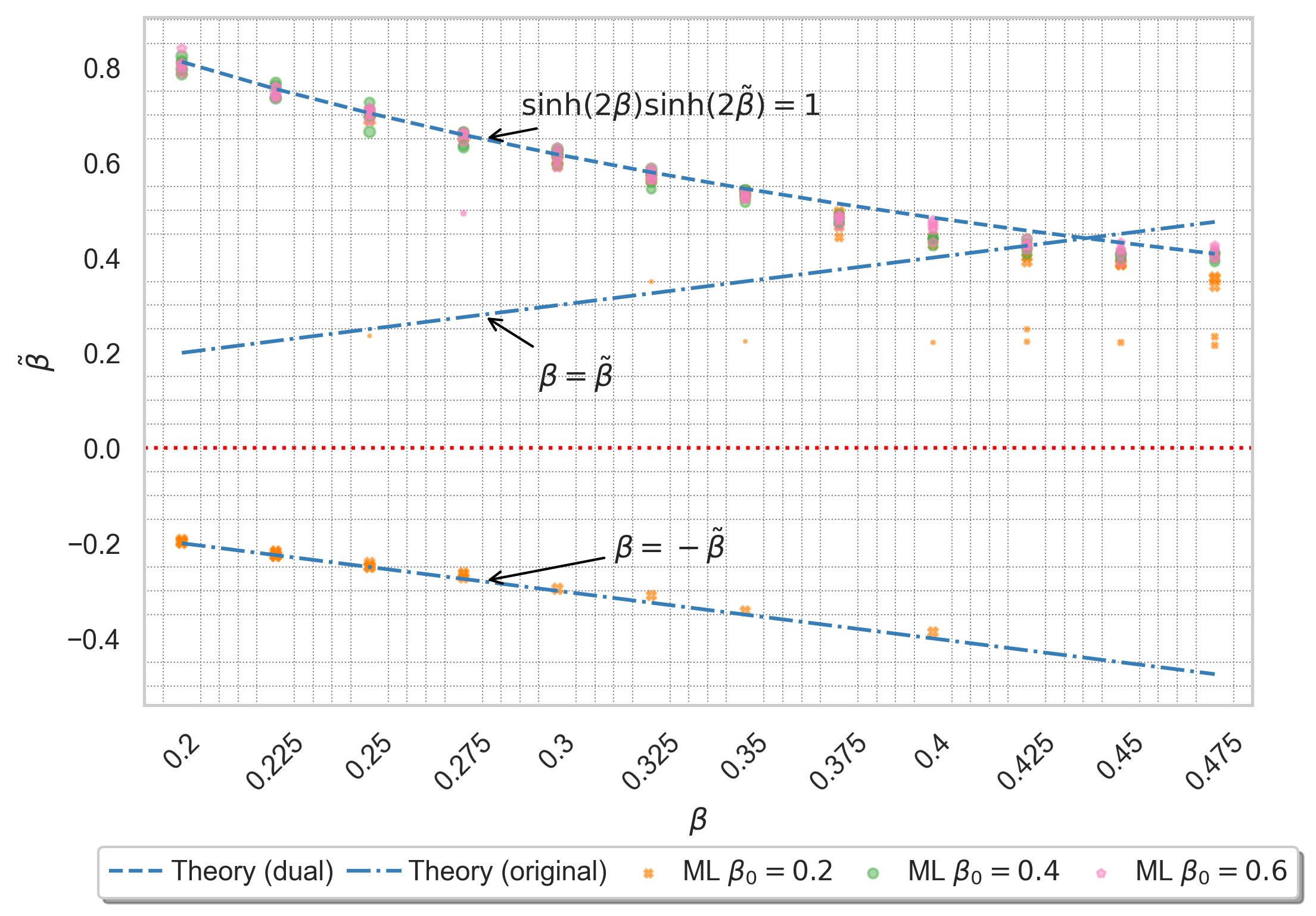}
    \caption{Final $\tilde\beta$ as found by the deep learning framework closely matches that of the theoretical results. Points are scaled by the negative logarithm of the best loss such that the size of the points is inversely proportional to the loss. We cap the minimum size so that smaller points are visible. The loss is a minimum along two fronts, i.e, original frame $\beta = \pm \tilde\beta$ and the dual frame along the lines $\sinh(2\beta)\sinh(2\tilde\beta) = 1$.   
    % \NI{I feel we need to explain the points that remain stuck at the starting point. Can we just exclude all points where the early stopping stops them before 1000 steps? Also what is the difference between orange and pink dots?}
    }.
    \label{fig:beta-betaTilde}
\end{figure*}
%\vspace{-1cm}
%\end{wrapfigure}

In this section, we describe some simple experiments using the above machinery. We take our original Hamiltonian $H$ to be that of the 2d Ising model \eqref{isingH}, and 
we take the dual Hamiltonian $\tilde{H}$ in~\eqref{dualH} to have only one non-zero parameter $\tilde{\beta}$ (and so $\tilde{\beta}_4=0$, etc.).

\paragraph{Neural Network architecture for $G_{\theta}$:} 
For a given link product in the dual frame we assemble the $7$ nearby links shown in  Fig. \ref{fig:linkMap} into a 7-dimensional vector $\mathbf{f}_{\vev{ij}} \in (\mathbb{Z}_2)^{7}$, where each element of the vector is the product of the two spins living on the two ends of the link. We consider a simplistic neural network acting on this input, with parameters formed by $\theta_1 \in \mathbb{R}^7$, and scalars $\theta_2$ and $\theta_3$. We opt for hard attention using Gumbel-Softmax~\cite{jang2016categorical} so that only a few of the seven nearby links are utilized in the prediction task.
%This constraint can also be enforced by augmenting the loss function with entropy minimization term over $\theta_1$.
%In this work, we chose to use the Gumbel-Softmax trick commonly used in the deep learning literature to sample discrete categorical variables in a differentiable manner. 
Thus, the mapping is defined by, 
\be
G_{\theta} (\mathbf{f}_{\vev{ij}}) = \theta_2 \cdot \text{Gumbel-Softmax}(\theta_1)^{T}\mathbf{f}_{\vev{ij}} + \theta_3
\label{eq:nnet}
\ee
As the elements of $\mathbf{f}_{\vev{ij}}$ are $\pm 1$, a very simple network provides a very expressive function. 
% \NI{Should we remove this paragraph for the short paper?} 
% We ran an ablation to check what would happen if we were not to use $\text{Gumbel-Softmax}$. We observe that a mapping of the form $G_{\theta} (\mathbf{f}_{\vev{ij}}) = \theta_2 \cdot \theta_1^{T}\mathbf{f}_{\vev{ij}} + \theta_3$ could work in some cases.
% Specifically, we see that while most of our training runs didn't yield a sensible outcome, some combination of features can yield an original frame or dual frame. 
% We show some of them in Figure~\ref{fig:no-gumbel-training}.

\paragraph{Rediscovery of the 2d Ising duality.}
In Figure~\ref{fig:beta-betaTilde}, we show the result of deploying the above machinery on different model values of $\beta$ on an $8 \times 8$ lattice with periodic boundary conditions. For each value of the input $\beta$, we ran a total of 15 optimizations, five from each of three initializations of $\tilde{\beta}$, i.e., $\tilde{\beta}_0=0.2$, $\tilde{\beta}_0=0.4$ and $\tilde{\beta}_0=0.6$. 
Due to the randomness involved in MCMC sampling, each seed is expected to be an independent run.

We record the value of $\tilde{\beta}$ obtained. There are three branches of solutions: the original model $\tilde{\beta} = \beta$, the dual model $\sinh(2\beta) \sinh(2\tilde{\beta}) = 1$, and an antiferromagnetic analogue of the original model $\tilde{\beta} = -\beta$. The latter is equivalent to the original duality frame, and is obtained by making the change of variables $\sig_i \to -\sig_i$ on every other site, thus flipping the sign of $\beta \to -\beta$. Note that the existence of the dual branch of solutions can be viewed as a numerical ``rediscovery'' of the KW duality line 
\be
\sinh(2\beta) \sinh(2\tilde{\beta}) = 1 \label{dualmap} 
\ee 
Further details on the experiments (including an exploration on how they depend on the system size) are shown in in Appendix \ref{app:exp}.

\begin{wrapfigure}[16]{r}{0.4\textwidth}
    \centering
    %\vspace{-0.75cm} 
    \includegraphics[width=0.95\linewidth]{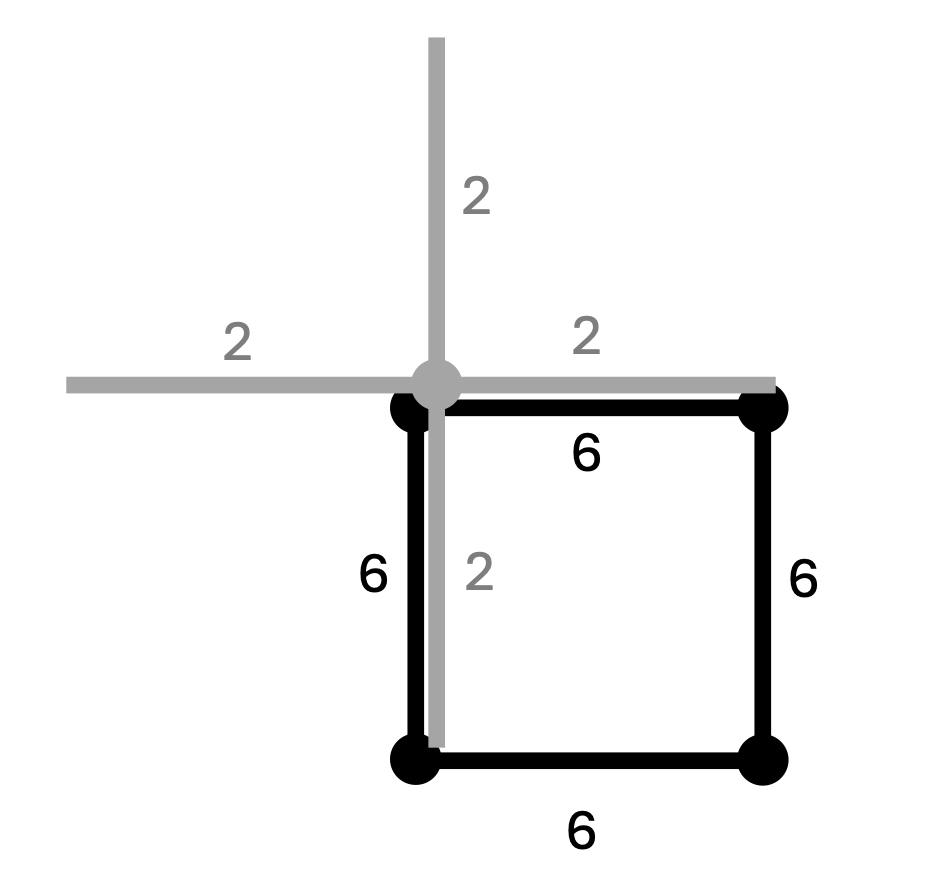}
    \caption{Emergence of dual lattice: e.g. if four original links (marked by $6$) form a square, the corresponding four links that are referenced by the neighbour mapping (marked by $2$) in Figure \ref{fig:linkMap} form a cross, as expected for the dual lattice.}
    %\NI{explain rot}   
    \label{fig:latticedual}
\end{wrapfigure}

As anticipated, it is interesting to ask how the model recovers the structure of the dual {\it lattice}, as well as the dual observables. The attention mechanism used encourages the model to use only a single link of the input, and for the runs that find the dual temperature this ends up using the links numbered either $2$ or $5$ instead of the original $6$ in Figure \ref{fig:linkMap}. As we show in an example in Figure \ref{fig:latticedual}, this is equivalent to finding the dual lattice from the original. 
 \label{sec:framework} Note that here it is important that we relate horizontal to vertical links by the composition of a rotation {\it and} reflection as shown in Figure \ref{fig:linkMap}; other choices will not result in the possibility of finding the dual lattice, and indeed in our experiments they do not find a duality. The optimized values of $G_\theta$ closely match theoretical results $G_{ij} (\tilde{\sigma}) = e^{-2\tilde{\beta} \tilde{\sig}_{i*} \tilde{\sig}_{j*}}$, as shown in more detail in Appendix \ref{app:exp}.

\section{Constructing the mapping using topological lines}\label{sec:top-lines}

The approach outlined in the previous section is very general, and the rather minimal ingredients used raise the exciting possibility of using the approach to search for completely new kinds of duality. 

However, in concrete examples of duality that we understand -- in particular those which are of the Kramers-Wannier type -- the realization of global symmetries plays a crucial role that has not been fully exploited in the framework above. In this section we outline a different approach, one where we fully use the structure of the global symmetries, showing that this dramatically reduces the complexity of the numerical problem, by essentially reducing it to a modified version of the inverse Ising problem. This of course comes at the cost of restricting us to duality of the order/disorder (i.e. Kramers-Wannier) type.

\subsection{Background: topological lines}
Let us begin by understanding a basic property of the spin correlations in the ordinary 2d Ising model, as reviewed in Section \ref{sec:dualitybg}. Consider a product of neighbouring spins $\sig_{i} \sig_{j}$ where $i_{i}$ and $i_{j}$ are connected by a single link. Let us take a product of $n$ such two-spin products $\sig_{i_1} \sig_{j_1} \sig_{i_2} \sig_{j_2}\cdots\sig_{i_n} \sig_{j_n}$ laid end-to-end along a continous path $C$ in an overlapping fashion, i.e. where $j_{n} = i_{n+1}$. At all intermediate sites along this path the spins themselves cancel, since $\sig_{i}^2 = 1$ for any site $i$. Thus this product is actually independent of the path $C$ itself, and only depends on the endpoints of the path:
\be
\prod_{\vev{ij} \in C} \sig_{i} \sig_{j} = \sig_{i_1} \sig_{j_1} \sig_{i_2} \sig_{j_2}\cdots\sig_{i_n} \sig_{j_n} = \sig_{i_1} \sig_{j_n} \label{path-indep}  \ . 
\ee
Now let us consider how this could be implemented in a duality. We seek a representation in terms of some dual spins $\tilde{\sig}_{\tilde{i}}$ with associated Hamiltonian $\tilde{H}$, where there is some function $G$ such that
\be
\vev{\sig_{i} \sig_{j}}_{H} = \vev{G(\{\tilde{\sig}\})}_{\tilde{H}} \ .  \label{tildeG2} 
\ee
The path-independence of the product of the bilinears $\sig_{i} \sig_{j}$ that is exhibited in \eqref{path-indep} is now very constraining: it tells us that an appropriate product of the $G$'s must {\it also} be path-independent. How can this be? 

This is implemented in Kramers-Wannier duality by realizing $G$ above as a link of a {\it topological line} which follows from the existence of a $\mathbb{Z}_2$ global symmetry. Indeed, whenever a system has a global symmetry, one can construct a {\it topological operator} -- in this case a topological line -- from it. This can be viewed as a generalization of Noether's theorem, and the rephrasing in terms of topological operators is most precise in the language of {\it generalized global symmetries} \cite{Gaiotto:2014kfa}, which has led to much recent progress in understanding quantum many-body systems (see \cite{Bhardwaj:2023kri,McGreevy:2022oyu,Shao:2023gho,Iqbal:2024pee} for reviews). 

We briefly review why the topological line exists in this specific context\footnote{See Appendix A of \cite{Iqbal:2020msy} for a more detailed discussion in similar language.}. Consider taking the dual Ising partition function \eqref{dualH}
\be
\tilde{H}[\tilde{\sig}] = -\tilde{\beta} \sum_{\vev{\tilde{i}\tilde{j}}} \tilde{\sig}_{\tilde{i}} \tilde{\sig}_{\tilde{j}} \qquad \tilde{Z} = \sum_{\tilde{\sig}_{\tilde{i}}} \exp(-\tilde{H})
\ee
and performing the change of variables $\tilde{\sig}_{\tilde{i}} = -\tilde{\sig}_{\tilde{i}}$ on a set of spins $i$ in some compact region $R$. As this is a change of variables, it leaves the partition sum $\tilde{Z}$ invariant. As it is a symmetry of the Hamiltonian, it leaves invariant all terms in the Hamiltonian that are either entirely inside $R$ or entirely outside $R$; however it has the effect of flipping the sign of the coupling $\tilde{\beta}$ for all terms that are on the boundary $C = \p R$. Thus this can be interpreted as inserting a closed line operator on $C$ into the partition function:
\be
\tilde{Z} = \sum_{\tilde{\sig}_{\tilde{i}}} \exp(-\tilde{H}) \prod_{\vev{\tilde{i}\tilde{j}} \in C} e^{-2\tilde{\beta} \tilde{\sig}_{\tilde{i}} \tilde{\sig}_{\tilde{j}}} = \bigg\langle\prod_{\vev{\tilde{i}\tilde{j}} \in C} e^{-2\tilde{\beta} \tilde{\sig}_{\tilde{i}} \tilde{\sig}_{\tilde{j}}}\bigg\rangle \label{invtop} 
\ee
where the factor of $2$ in the exponent arises from the need to flip the sign of each term multiplying $\tilde{\beta}$. 

Clearly the precise boundary of this region $R$ does not matter. Thus we see that in this Hamiltonian, a product of links of the form $e^{-2\tilde{\beta} \tilde{\sig}_{\tilde{i}} \tilde{\sig}_{\tilde{j}}}$ forms a topological line, in that its expectation value is invariant under deformations of $C = \p R$. 

In Kramers-Wannier duality, this construction is how the path-independence exhibited in \eqref{path-indep} is implemented. The product of spin bilinears $\sig_i\sig_j$ is dual to a chunk of the topological line:
\be
\sig_i\sig_j = e^{-2\tilde{\beta} \tilde{\sig}_i* \tilde{\sig}_j*} \label{topdualrep} 
\ee
where here the notation $i^*$ and $j^*$ is as shown in Figure \ref{fig:duallink}. The purely kinematical path-independence of \eqref{path-indep} is replaced by the path independence associated with the global symmetry as described above. 

We now see that (in the absence of any other topological lines in the theory) the mapping could not have taken any other possible form. 

We also see that exploiting this global symmetry gives us a concrete way to proceed and search for dualities. Given a more general dual Hamiltonian with a set of local couplings $\{ \tilde{\beta}_{\tilde{i}} \}$, we can explicitly construct the form of the topological line by examining the transformation of the Hamiltonian under the global symmetry $\sig \to -\sig$. In this way will {\it directly} find a candidate function $G(\tilde{\sig}_{\tilde{i}})$ as in \eqref{tildeG2} which manifestly satisfies the path-independence shown in \eqref{path-indep}, a requirement for it to be dual to $\sig_i \sig_j$. Thus, this reduces the problem of the previous part to simply determining the couplings of the dual Hamiltonian, a much simpler numerical problem. 
\subsection{Methods}
We now describe how this approach is implemented. In this section we will work with more general Hamiltonians. 

\subsubsection{Constructing topological line}
We first have to describe how to construct the topological line described above for a general $\mathbb{Z}_2$-invariant Hamiltonian. This is a problem in bookkeeping. Consider a Hamiltonian density $\mathcal{H}$ of the general form
\be
\mathcal{H}_i = -\sum_{a} \beta_a N_{a,i}[\sig]
\ee
where the sum $a$ runs over independent terms of the Hamiltonian, each $\beta_{a}$ is a coupling, and where each $N_{a,i}$ is a local product of an even number of spins centered around the site $i$. By ``local'' we mean that this can be constructed by taking products of different spins which are connected by only a single link, i.e. if we denote this set of links by $L_{a}$ we have
\be
N_{a,i}[\sig] = \prod_{\vev{jk} \in L_{a}} \sig_{j} \sig_{k}
\ee
For example, for the usual Ising model, there are two $L_{a}$, corresponding to horizontal and vertical links respectively. Each of them contains only one link, which we express in the following relative fashion oriented around the site $i = (x,y)$: 
\be
L_{\mathrm{horizontal}} = \{(x,y) \leftrightarrow (x+1,y) \} \qquad L_{\mathrm{vertical}} = \{(x,y) \leftrightarrow (x,y+1) \}
\ee
and by rotational invariance the corresponding couplings are the same $\beta_{\mathrm{horizontal}} = \beta_{\mathrm{vertical}} = \beta$. 
The full Hamiltonian is constructed by summing the density $\mathcal{\sH}_i$ over all sites in the lattice, $H[\sig] = \sum_{i} \mathcal{\sH}_i[\sig]$.  

An efficient way to do the book-keeping for constructing the topological line is to gauge the $\mathbb{Z}_2$ symmetry, i.e. to introduce an {\it external} $\mathbb{Z}_2$-valued gauge field $t_{\vev{ij}}$ on the links of the lattice, and to replace in the Hamiltonian each product of spins connected by a single link with the following {\it gauged} version:
\be
\sig_{i} \sig_{j} \to \sig_{i} t_{\vev{ij}} \sig_{j}
\ee
The new Hamiltonian $H[\sig;t]$ constructed in this manner is now invariant under the following operation for an arbitrary $\mathbb{Z}_2$ valued function $\lam_i$ of the sites:
\be
\sig_{i} \to \sig_{i}' = \lam_{i} \sig_{i} \qquad t_{\vev{ij}} \to t_{\vev{ij}}' = \lam_{i} t_{\vev{ij}} \lam_{j} \label{trans} 
\ee
and thus the partition function -- which is now a function of the external $t_{\vev{ij}}$ -- is invariant under the same transformation, which acts now only on $t$: 
\be
Z[t] = \sum_{\sig_i} \exp(-H[\sig;t]) = Z[t']
\ee
This invariance is equivalent to the arguments around \eqref{invtop} and allows us to construct a topological line. 

More explicitly, consider drawing a path $C$ on the dual lattice which intersects a set of links of the original lattice. Set $t_{\vev{ij}} = -1$ on all links intersected by the path and $t_{\vev{ij}} = +1$ on all the rest of the links. This defines a line operator on the curve $C$. Denote the partition function in the presence of this source by $Z[t(C)]$. By using transformations of the form \eqref{trans} the pattern of nontrivial $t_{\vev{ij}}$ can be altered in such a manner that the curve $C$ is continuously deformed without altering the partition function, thus illustrating that the line $C$ is topological. 

We can express this line operator as an expectation value in the original Hamiltonian. Note that setting $t_{\vev{ij}} = -1$ along a curve $C$ has the effect of flipping the signs of some of the $\beta_{a}$ in the expansion. Given an explicit parametrization of the curve $C$, it is a book-keeping problem to keep track of which $\beta_{a}$ on which sites $i$ are so modified and then write the partition function as
\be
Z[t(C)] = \sum_{\sig_i} \exp(-H[\sigma]) \prod_{i,a \in S} \exp(-2\beta_{a} N_{a,i}[\sig]) \label{Cdef} 
\ee
for an appropriate set $S$. 

\subsubsection{Matching to dual Hamiltonian} 
We now use this technology to find pairs of Hamiltonians related by duality. In addition to a ``primal'' Hamiltonian $H$ defined as above, consider also a {\it dual} Hamiltonian with couplings $\tilde{\beta}_a$ and spins denoted by $\tilde{\sig}_{i^*}$. We denote points on the lattice where the dual $\tilde{\sig}$ live by starred coordinates $i^*$.  
\be
H = - \sum_{i} \sum_{a} \beta_a N_{a,i}[\tilde{\sig}] \qquad 
\tilde{H} = - \sum_{i^*} \sum_{a} \tilde{\beta}_a N_{a,i^*}[\tilde{\sig}] 
\ee
We apply the topological line approach to the dual Hamiltonian, and construct the operator associated with a curve $C$ that is only a single link $\vev{i^*j^*}$; we denote this operator as $G_{\vev{i^*j^*}}$, and it takes the form
\be
G_{\vev{i^*j^*}} = \prod_{k^*,a \in S_{\vev{i^*,j^*}}}\exp(-2\tilde{\beta}_{a} N_{a,k^*}[\tilde{\sig}]) 
\ee
where the set $S$ of the couplings that we flip depends implicitly on the link chosen. Now by the arguments around \eqref{topdualrep}, the Hamiltonian $H$ is a Kramers-Wannier dual to $\tilde{H}$ if the following relation holds: 
\be
\langle \sig_i \sig_j \cdots \rangle_{H} = \bigg\langle \prod_{k^*,a \in S_{i^*j^*}}\exp(-2\tilde{\beta}_{a} N_{a,k^*}[\tilde{\sig}]) \cdots \bigg\rangle_{\tilde{H}} \ , \label{topdualrel} 
\ee
i.e. if the link of a topological line can be realized directly in terms of fundamental spins in the dual frame. 

In this section, we will assume that we are given the {\it dual} couplings $\tilde{\beta}_a$, and our task is to find the couplings in the original frame $\beta_a$ such that \eqref{topdualrel} is satisfied. This is a significantly easier problem than that of the previous section, because we used physics input to completely determine the form of the function $G$; we thus need only determine the couplings. By taking appropriate products of the $G_{\vev{i^* j^*}}$ and doing Monte Carlo simulations using $\tilde{H}$ we can compute arbitrary moments of the right-hand side of \eqref{topdualrel}. 

Thus the problem is equivalent to determining the couplings of a generalized Ising model $H[\sigma]$ given all moments $\vev{\sig_i \sig_j \cdots}_H$. This problem is sometimes called the ``inverse Ising problem'', with a large literature: see \cite{nguyen2017inverse} for a review of different approaches. We will use an algorithm from  \cite{PhysRevLett.52.1165,aurell2012inverse} which essentially exploits a self-consistency relation on correlation functions. 

 \subsubsection{Numerical implementation} 

We briefly review the algorithm, following the implementation outlined in \cite{albert2014inverse}.  For the remainder of this section we will consider only generalized Ising models with two-spin interactions; thus the Hamiltonian takes the form
\be
H = -\sum_{i} \sum_{\delta} K_{i,i+\delta} \sig_i \sig_{i+\delta}
\ee
where $K_{i,i+\delta}$ is a 2-site coupling, and $\delta$ should be understood as a displacement vector taking us from a site $i$ to a nearby (but not necessarily connected by a single link) site $i+\delta$. We can derive a useful consistency relation by noting that the marginal probability distribution of a given $\sig_i$ depends only on its neighbours; explicitly summing over a single $\sig_i$ we find
\be
\vev{\sig_i} = \tanh\le(\sum_{i,\delta} K_{i,i+\delta} \sig_{i+\delta}\ri)
\ee
which is the analogue of an ``equation of motion'' in this discrete statistical mechanics setting. In a $\mathbb{Z}_2$ symmetric setting on a finite lattice both sides of the above equation identically vanish, but the following generalization is useful: 
\be
\vev{\sig_j \sig_{j+\delta}} = \bigg{\langle} \tanh\le(\sum_{\delta'} K_{j,j+\delta'}\sig_{j+\delta'}\ri) \sig_{j+\delta}\bigg{\rangle} \label{Krel} 
\ee
The idea of the algorithm is simply that given a set of samples of the $\{\sig_i \}$ from Monte Carlo simulations, we adjust the $K_{i,i+\delta}$ so that \eqref{Krel} holds. 

At a practical level, we do this by computing the following gradient 
\be
D_{\delta,\delta'} \equiv \frac{\p}{\p K_{j,j+\delta'}} \vev{\sig_j \sig_{j+\delta}} = \bigg{\langle}\sig_{j+\delta'}\sig_{j+\delta}\; \mbox{sech}^2\le(\sum_{\delta''} K_{j,j+\delta''}\sig_{j+\delta''}\ri)\bigg{\rangle}
\ee
Note that in a situation with translational invariance, $K_{i,i+\delta} = K_{\delta}$ alone, and similarly all correlation functions above do not depend on the reference point $j$, as the notation for $D_{\delta,\delta'}$ implicitly assumes. 

We then implement a variant of Newton's method for $K_{\delta}$ by solving the linearized equation:
\be
\vev{\sig_j \sig_{j+\delta}} - \bigg{\langle} \tanh\le(\sum_{\delta'} K_{\delta'}\sig_{j+\delta'}\ri) \sig_{j+\delta}\bigg{\rangle} = \sum_{\delta'} D_{\delta, \delta'} \Delta K_{\delta'} \label{Newton} 
\ee
for the increment $\Delta K_{\delta'}$ and iterating as $K_{\delta} \leftarrow K_{\delta} + \ep \Delta K_{\delta}$ for some learning rate $\ep$. The fixed point of this procedure will drive the left-hand side to zero and thus find the values of the couplings $K$ that are consistent with the given data.  

We now come to an interesting point. Note that the input into this procedure requires us to evaluate non-linear correlation functions of the $\sig_j$; in usual applications such as those discussed in \cite{PhysRevLett.52.1165,aurell2012inverse,albert2014inverse} one is given the $\sig_i$ directly in the form of samples of MCMC data and it is straightforward to estimate the correlation functions required from the samples. In our case, however, we are not directly given the $\sig_i$; instead all we can do is extract  $\mathbb{Z}_2$ even correlation functions of the $\sig_i$, in terms of correlation functions of the dual spins $\{\tilde{\sig}_{\tilde{i}}\}$ through relations such as \eqref{topdualrel}. As all $\mathbb{Z}_2$ odd correlation functions vanish above in a finite system, in principle this is the required information, and we could attempt to extract it by Taylor expanding various functions as  $\tanh(x) = x + \cdots$. These expansions will eventually truncate at some order since $\sig_i^2 = 1$ for each spin. In practice however the combinatorics involved in this is intractable.  

We will thus extract it in a different manner. Due to the locality of the couplings $K_{\delta}$, the data required is the marginal distribution of the spins in a $N \times N$ window, where $N$ is twice the range of the interaction $K_{\delta}$. If this $N$ is small enough we can now perform exact summation over these degrees of freedom. More explicitly, let us denote these spins as $\sig_a$; their marginal distribution $p(\{\sig_a\})$ can be thought of a $2^{N^2}$ dimensional vector. The following relation is the definition of the expectation value:
\be
\sum_{\{\sig_a \}} p(\{\sig_a\})(\sig_{1})^{\alpha_1} (\sig_{2})^{\alpha_2} \cdots (\sig_{N^2})^{\alpha_{N^2}} = \vev{(\sig_{1})^{\alpha_1} (\sig_{2})^{\alpha_2} \cdots (\sig_{N^2})^{\alpha_{N^2}}}  \label{pdual} 
\ee
where each of the $\alpha_{a} \in \{0, 1\}$ determines whether the corresponding spin $\sig_a$ appears in the expectation value or not. As there are $2^{N^2}$ choices for the $\alpha_a$, this can be viewed as a matrix equation for $p(\{\sig_a\})$ and thus we can in principle invert this matrix to find the marginal distribution $p(\{\sig_a\})$. This is only practical if $N$ is small enough. 

Finally, we use this distribution to evaluate all non-linear correlation functions appearing in \eqref{Newton} as
\be
\vev{f(\{\sig_a \})}= \sum_{\{\sig_a \}} p(\{\sig_a\}) f(\{\sig_a\}) \label{nonl} 
\ee
for an arbitrarily complicated non-linear function $f$. 

To summarize, to obtain the Kramers-Wannier dual for a given Hamiltonian specified by couplings $\tilde{\beta}$ we do the following:
\ben
\item We decide on the couplings $K_{\delta}$ that we allow to be present in the putative dual and let $N$ be twice their range. 
\item We perform MCMC sampling of the Hamiltonian specified by $\tilde{\beta}$ and use \eqref{topdualrel} to compute all correlation functions of the putative dual spins $\sig_a$ in a $N \times N$ window. 
\item We insert these correlation functions into \eqref{pdual} to solve for the marginal distribution $p(\{\sig_a\})$ \ . 
\item 
 We  find the optimum $K_{\delta}$ by computing the gradient $\Delta K_{\delta}$ using \eqref{Newton} and iterating, where all nonlinear correlation functions are computed using \eqref{nonl}. 
 \een

\subsection{Dualities of nearest-neighbour couplings}

We now discuss a specific model, followed by the results from this approach.
\subsubsection{Next-to-nearest neighbour Ising model} 
\begin{wrapfigure}[16]{r}{0.3\textwidth}
    \centering
    %\vspace{-0.75cm} 
    \includegraphics[width=0.95\linewidth]{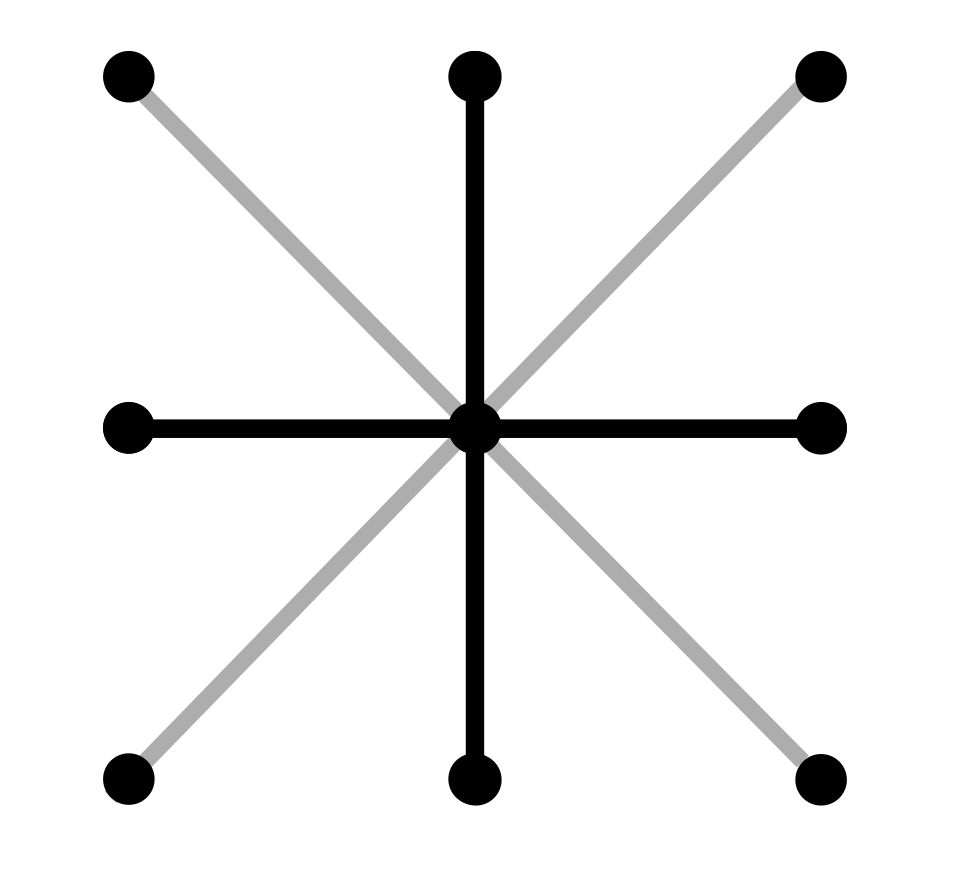}
    \caption{Links connecting a site to its neighbours; those in black are weighted by $\tilde{\beta}$ and those in grey by $\tilde{\kappa}$. }
    %\NI{explain rot}   
    \label{fig:8links}
\end{wrapfigure}

 We consider as a testbed the following the next-to-nearest neighbour (NNI) Ising model:
\be
\tilde{H}[\tilde{\sigma}] = -\tilde{\beta} \sum_{\vev{\tilde{i}\tilde{j}}} \tilde{\sig}_{\tilde{i}} \tilde{\sig}_{\tilde{j}} - \tilde{\kappa} \sum_{[\tilde{i}\tilde{j}]} \tilde{\sig}_{\tilde{i}} \tilde{\sig}_{\tilde{j}}
\ee \label{kappaham} 
The notation $\vev{\tilde{i}\tilde{j}}$ refers to the usual nearest-neighbour coupling -- i.e. sites differing by the vectors $(0,1)$ or $(1,0)$. Here $[\tilde{i}\tilde{j}]$ refers to a {\it diagonal} coupling, i.e. two sites $\tilde{i},\tilde{j}$ which differ by $(+1,+1)$ or $(+1,-1)$, and is thus a next-to-nearest-neighbour coupling, as shown in Figure \ref{fig:8links}.

% However, in the course of our investigation we came to learn that an exact Kramers-Wannier-type dual for this model was analytically constructed in 2014 \cite{strycharski2013dual}. Interestingly, this dual construction requires one extra spin per unit cell and would not be found by our current framework. We believe that the duals that we find may be understood as {\it approximations} to this exact duality arising from integrating out the extra degree of freedom. Integrating this out should generically result in mild non-localities, in that in a neighbourhood of the Ising point at $\ka = 0$ the exact dual should contain spin-spin couplings that decay exponentially with distance. We are at present unable to test this for technical reasons, but we comment on prospects for doing this in the conclusion.

\paragraph{Physics of the model and analytic dual.}

This model has been previously studied as an extension to the standard Ising model \cite{zandvliet20062d,lee2010study,strycharski2013dual}, and we briefly discuss its physics. At $\tilde{\ka} = 0$ this reduces to the usual Ising model with a phase transition at $\tilde{\beta} \approx 0.44$. At $\tilde{\beta} = 0$ it reduces to two disconnected Ising models, each living on one of the two checkerboard bipartite partitions of the square lattice; thus this model also has an ordering transition at $\tilde{\ka} \approx 0.44$, though there is now an extra global symmetry corresponding to flipping the spins on each checkerboard independently. Numerical investigations show that the phase transition line connects these two points, and one might expect the transition to remain in the usual 2d Ising class in the neighbourhood of $\tilde{\ka} \approx 0$. 

 An analytic dual for this model was implicitly constructed in~\cite{strycharski2013dual} using transfer matrix methods (building on~\cite{onsager1944two,schultz1964two}). Due to the non-planarity of the couplings -- i.e. the fact that the links cannot be drawn on the plane without crossing -- it is not immediately clear how the dual lattice where the dual model is defined has to look like. However, one can view the original lattice as a triangular lattice (given for instance by those links that are horizontal, vertical, and those that go from bottom-right to top-left in Figure~\ref{fig:8links}), with an addition of some non-planar coupling. The dual of a triangular lattice is a honeycomb lattice, and so it is not surprising that~\cite{strycharski2013dual} finds a dual model defined on a lattice of the same topology as the honeycomb lattice, but with in addition a quartic coupling of spins (which is argued to originate from the non-planarity of the original model). Interestingly one could also view this dual as a model with the same unit cell as a square lattice, but with twice as many spins (with a particular pattern of couplings) living on the unit cell as in the original model. 

\paragraph{Numerical exploration as a deformation of the 2d Ising duality.} Our motivation to study this Hamiltonian is that it is the simplest model that allows us to explore the robustness of the algorithm to turning on extra couplings from known dualities, and we will take a slightly different but still informative approach. In particular, we will consider this duality as a deformation of the standard 2d Ising duality discussed above, and attempt to find a candidate dual Hamiltonian of the {\it same} form (i.e. also labeled by two couplings $\beta$, $\kappa$) defined on a dual \emph{square} lattice. We note that this candidate dual Hamiltonian does not fit into the class of Hamiltonians found in \cite{strycharski2013dual}, and thus it should only be viewed as an approximate duality, i.e. an effective description of the exact dual to the NNI Ising model (found by perturbing around the original Ising model).  We wish to see how well this effective description can be brought to match correlation functions, and we will notice remarkable agreement.

\subsection{Results}
We now display the results of some preliminary investigations. As a check on the stability of the numerics we treat as independent parameters the coupling on each of the eight links emerging from the central spin as in Figure \ref{fig:8links}. The structure of the couplings that we allow means that the size of the window in question in \eqref{pdual} is $N = 3$. The matrix inversion in \eqref{pdual} is over only a $512 \times 512$ matrix and can easily be done. 

\begin{figure}[htbp]
    \centering
    % First subfigure
    \begin{subfigure}{0.45\textwidth}    % 0.32\textwidth for equal spacing in 3 figures
        \includegraphics[width=\linewidth]{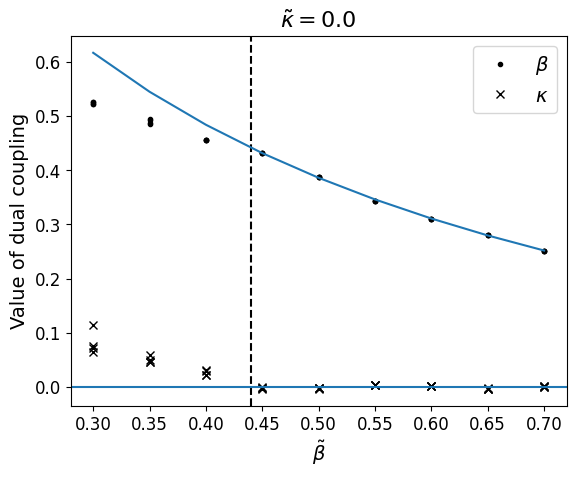}
        %\caption{Normal KW duality; blue lines note expected theoretical result, which we see is recovered for $\tilde{\beta} > \beta_{c}$.}
        \label{fig:enter-label}
    \end{subfigure}
    % Second subfigure
    \begin{subfigure}{0.45\textwidth}
        \includegraphics[width=\linewidth]{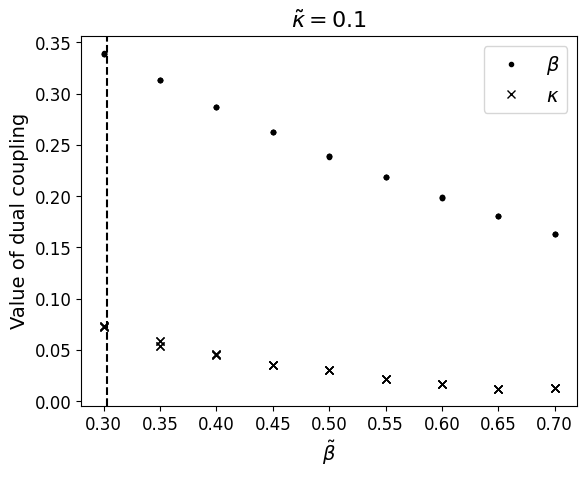}
        %\caption{Caption for image 2}
        \label{fig:subfig2}
    \end{subfigure}
    % Third subfigure
    %\begin{subfigure}{0.32\textwidth}
        %\includegraphics[width=\linewidth]{image3.png}
    %    \caption{Caption for image 3}
    %    \label{fig:subfig3}
    %\end{subfigure}
    
    \caption{Examples of determination of $\beta, \kappa$ as a function of $\tilde{\beta}, \tilde{\kappa}$. Location of phase transition $\tilde{\beta}_c$ is indicated with dashed vertical line; note that the algorithm only functions reliably for $\tilde{\beta} > \tilde{\beta}_c$. Left-hand panel is regular KW duality, and there we indicate theoretical expectations with solid blue line. }
    \label{fig:findingcouplings}
\end{figure}

Some representative results are shown in Figure \ref{fig:findingcouplings}. The case $\tilde{\ka} = 0$ is just conventional KW duality. We see that for $\tilde{\beta} > \tilde{\beta}_c$ -- i.e. when the dual is in the ordered phase -- we are able to reconstruct the known 2d Ising duality mapping numerically. Note that -- unlike in Section \ref{MLapproach}, where we only allowed for a single coupling $\beta$ -- here we allow for a nonzero dual value of $\kappa$, but it is driven to zero dynamically. We stress that the algorithm fails in the disordered phase $\tilde{\beta} < \tilde{\beta}_c$, and we can see that there the eight independent values for the couplings no longer cluster as expected into two sets of couplings. It is interesting that the algorithm fails in the region where we are attempting to learn parameters $(\beta, \kappa)$ for an ordered Ising model. This appears to be related to known issues associated with learning parameters of classical Hamiltonians at large $\beta$ (see e.g. Appendix B of \cite{Haah:2021nzn}) and deserves further study. 

\begin{figure}[htbp]
    \begin{subfigure}{0.45\textwidth}
        \centering
        \includegraphics[width=\linewidth]{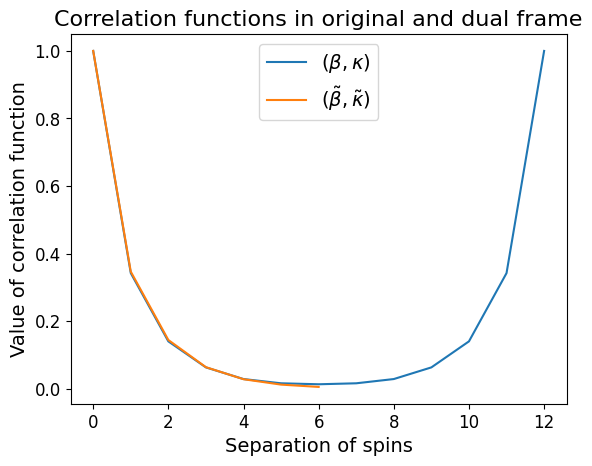}
        %\caption{Enter Caption}
        \label{fig:corrfunc}
    \end{subfigure}
    \begin{subfigure}{0.45\textwidth}
        \centering
        \includegraphics[width=\linewidth]{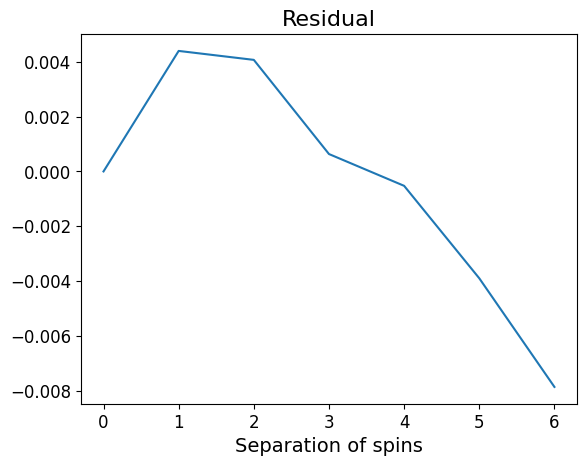}
        %\caption{Enter Caption}
        \label{fig:residuals}
    \end{subfigure}
    \caption{Spin-spin correlation functions measured in original and dual frame with $(\beta,\kappa) = (0.261,0.0194)$ , $(\tilde{\beta},\tilde{\kappa}) = (0.45,0.05)$. Left panel shows the correlation functions themselves; right panel shows the difference between these two correlation functions, demonstrating that it increases with separation.} 
    \label{fig:corr} 
\end{figure}

We now turn on a small nonzero $\tilde{\kappa}$. We see that the curve for $\beta(\tilde{\kappa}, \tilde{\beta})$ is modified slightly and that the value of $\kappa(\tilde{\kappa}, \tilde{\beta})$ is now nonzero, changing continuously as a function of $\tilde{\beta}$. This can be understood as a ``best fit'' dual given the restricted number of couplings that have been activated. To estimate the goodness of the fit, In Figure \ref{fig:corr} we compute the two-point correlation function $\vev{\sig_{i} \sig_{i+\delta}}$ on a $12 \times 12$ lattice in both frames, where in the frame defined by $(\tilde{\beta}, \tilde{\ka})$ it is computed using \eqref{topdualrel}. Note that the duality as we have formulated it is exact only in the thermodynamic limit, and is broken by topological excitations that wrap the two non-trivial cycles of the torus. This means that correlation functions that probe more than half of the system will certainly not agree -- in the case at hand the dual correlator will continue to decay exponentially and not respect the periodicity of the torus, as there is a topological line stretching between the two points. Thus we truncate the dual frame correlator at half of the system size. 

In this regime the agreement is impressive, and in some preliminary experiments it remains so as we vary $\tilde{\ka}$. Nevertheless, for the reasons explained above we do not expect this to be an exact dual and expect that the performance should eventually degrade, possibly as we move further from the Ising point at $\tilde{\ka} = 0$ or as we go to larger system sizes allowing the probing of longer correlation functions. We compute the small difference between the correlation functions in the two frames in the right-hand panel of Figure \ref{fig:corr}, indeed demonstrating that the agreement becomes worse at larger separations. 

\section{Conclusions and outlook} 

In this work, we have formalized the task of finding dualities in classical statistical physics in such a way that it can be tackled by machine learning. The key insight is that in order to find dual candidates of a given model, one needs to generalize the classical problem of learning couplings $\tilde{\beta}$ in a Hamiltonian $\tilde{H}[\tilde{\beta} , \tilde{\sig}]$ starting from data to the simultaneous learning of
\begin{enumerate}
\item the couplings $\tilde{\beta}$,
\item some properties of the underlying space $\tilde{L}$ in which the model is defined,
\item a mapping of observables $G(\tilde{\sig})$.
\end{enumerate}
In Section \ref{MLapproach} -- which we call the ``machine learning approach'' -- we have introduced a loss that measures the distance between moments of observables across the (given) original model and candidate dual models, and we have discussed algorithms aimed at minimizing it. We have demonstrated that this optimization problem is well-posed in the prototypical example of the 2d Ising model, and that its solution leads to a genuine automated ``rediscovery`" of the celebrated Kramers-Wannier duality of the 2d Ising model.

In Section \ref{sec:top-lines} -- the ``topological lines approach'' -- we we have discussed a different numerical method that can exclusively be applied to the case of dualities of Kramers-Wannier type in lattice models. Assuming that the dual model is defined on the dual lattice, symmetry-based arguments can be used to explicitly construct the form of the function $G(\tilde{\sig})$ in terms of the couplings in the dual Hamiltonian $\tilde{H}[\tilde{\beta},\tilde{\sigma}] $. The only outstanding task is then to optimize the couplings $\tilde{\beta}$ to find agreement between the original and dual models. In this restricted context, it is possible to efficiently solve this optimization problem by importing algorithms developed in the context of the inverse Ising problem, thereby producing a resilient method to find and explore dualities. We once again applied this technology to the 2d Ising model, and investigated some mild deformations by adding next-to-nearest neighbour couplings, finding a putative effective dual that displays impressive agreement of correlation functions.

We note that in all of our approaches the relation between original and dual couplings $\beta$ and $\tilde{\beta}$ is only found numerically; one could possibly supplement this numerical determination with symbolic regression~\cite{schmidt2009distilling} to obtain analytic formulas such as \eqref{dualmap}. In more complicated examples of duality we do not expect there to necessarily exist analytic relations and thus have not explored this here.

In this work we have demonstrated the feasibility of such approaches by reconstructing known results and considering deformations thereof. In the future we plan to use this approach to systematically search for {\it new} dualities. There are some challenges to be overcome in this process, most of which are (in our opinion) technical:
\ben
\item  In the machine learning approach, the loss function we have introduced is natural and easily generalizable, but it would be desirable to have a more universal loss function whose minima are guaranteed to be exact dualities. This would likely require a generalization of the kernel method to a set-up where only appropriate moments are guaranteed to be matched, and where the matching of moments carrying little information (for instance, measuring correlations of far-away spins) is suppressed.

\item  Similarly, computing gradients in the machine learning approach is currently very expensive. This is because, unlike in the classical case of Boltzmann machines, we cannot use techniques such as contrastive divergence, and we are forced to fully equilibrate an MCMC chain in each step. It would be very useful to find more efficient methods to compute the gradients.
\item Initial explorations show that with the addition of more couplings in the machine learning approach, training starts to become unstable. An improved optimization algorithm would therefore be desirable. It is possible that if an appropriate kernel method is identified, then the training will become more stable.
\item The ``topological lines'' approach is more conservative, as it will only find order/disorder dualities which are of the Kramers-Wannier type, and thus likely continuously connected to known dualities. Relatedly, most steps of the the ``topological lines'' approach are computationally inexpensive and can readily be generalized to more complicated systems. The exception is a single step, described below \eqref{Newton}, where non-linear correlation functions in the dual frame must be computed from correlation functions in the original frame. At present this is done by computing an exact marginal probability distribution by inverting a matrix in \eqref{pdual}, one whose size grows exponentially with the range of the local couplings in the model. This will rapidly become infeasible if this range is increased. This is the primary obstacle to probing longer range couplings. This is important as we would like to increase the range to determine whether any candidate duals are truly local, or to make contact with the results of \cite{strycharski2013dual}. Ideally one would like to {\it allow} for couplings at longer ranges and verify that they are dynamically driven to zero.

This approach is a brute-force one that makes no use of the structure present and we imagine one could obtain the information in some other way, e.g. through training a miniature neural network to obtain the required correlation functions. 
\een

We further note that the only topological lines that we have considered are those associated to the $\mathbb{Z}_2$ spin-flip symmetry. Recently it has been fruitful to interpret Kramers-Wannier duality in terms of a non-invertible symmetry \cite{Frohlich:2004ef,Chang:2018iay}, i.e. a topological line which implements the action of the duality at the self dual point $\beta = \tilde{\beta}$  (see \cite{Shao:2023gho} for a review from different points of view). It would be interesting to consider the interplay of these ideas, e.g. by searching for new microscopic self-dual lattice systems, or by attempting to use numerical optimization tools such as those studied in this work to systematically search for new kinds of emergent topological line.

We will return to these issues in the future. A concrete target is to map out the space of dualities of deformations of the classical 2d Ising model. A more ambitious goal is to utilize such an approach to probe new and unexpected kinds of duality in statistical physics. 

\vspace{1cm}
{\bf Acknowledgments:} We are very grateful to Roberto Bondesan, Arkya Chaterjee, Tarun Grover, Tyler Helmuth, Theo Jacobson, John McGreevy, Takuo Matsubara, Salvatore Pace and Tin Sulejmanpasic for helpful discussions. NI is supported by a Simons Pivot Fellowship and in part by the STFC under grant number ST/T000708/1. AEVF was in part supported by the EPSRC Grant EP/W020939/1 ``3d N=4 TQFTs". This work has made use of the Hamilton HPC Service of Durham University.

\begin{appendix}
\section{Kramers-Wannier duality} \label{app:KW}
For completeness, here we present a derivation of the usual Kramers-Wannier duality for the 2d Ising model on the square lattice\footnote{We are grateful to T. Sulejmanpasic and T. Jacobson for discussions on the content of this section.} in a manner that shows how the duality is complicated when the model is deformed away from the standard nearest-neighbour couplings. 

\subsection{Usual Kramers-Wannier duality} 
Consider the usual classical 2d Ising model with spins $\sig_i = \pm 1$ living on the sites of a 2d lattice. 
\be
H[\sig] = -\beta \sum_{\vev{ij}} \sig_i \sig_j \qquad Z = \sum_{\{\sig_i\}} \exp(-H) \label{ungauged} 
\ee
Here we have chosen to absorb the inverse temperature $\beta$ into the Hamiltonian. The notation $\vev{ij}$ refers to the link connecting nearest neighbour sites $i$ and $j$. The model is invariant under the global $\mathbb{Z}_2$ symmetry which acts simultaneously on all spins as $\sig_i \to -\sig_i$. 

To derive the duality, we will consider a {\it gauged} version of the model.    In other words, we consider the following model with an extra dynamical $\mathbb{Z}_2$-valued gauge field $t_{ij}$ living on the links. 
\be
H'[t,\sig] = -\beta \sum_{\vev{ij}} t_{ij} \sig_i \sig_j \qquad Z' = \sum_{\{\mathrm{flat}\;t_{ij}\}}\sum_{\{\sig_i\}} \exp(-H) \label{gaugedZ2def} 
\ee
This model has an enlarged $\mathbb{Z}_2$ gauge symmetry which acts as
\be
t_{ij} \to \lambda_i t_{ij} \lambda_j \qquad \sig_i \to \lambda_i\sig_i
\ee
where $\lambda_i$ is an arbitrary $\mathbb{Z}_2$-valued function living on the sites. 

The injunction to sum only over {\it flat} $t_{ij}$ means that we include in the partition sum only $t_{ij}$ that have vanishing field strength, i.e. if we take the product over all the links $l$ of a plaquette $p$ to define a plaquette-valued field strength $F_{p} \equiv \prod_{l \in p} t_l$ we find $1$ for {\it all} plaquettes:
\be
F_{p} = 1 \qquad \leftrightarrow \qquad t_{ij}\;\mbox{flat} \label{flatness} 
\ee
It is well-known that gauging a global symmetry in this flat manner results in a theory with essentially the same {\it local} physics in the symmetry-invariant sector, though global issues (i.e. involving non-local operators, or the behavior of manifolds on different topology) can differ. We will explicitly demonstrate this fact and use it to compute the partition sum $Z'$ in two different ways. 
\ben
\item
In the first, we note that since $t_{ij}$ is flat it can locally be written in the ``pure gauge'' form as $t_{ij} = \lam_{i} \lam_{j}$ with $\lam_{i}$ a $\mathbb{Z}_2$-valued function on the sites. This manifestly satisfies the condition \eqref{flatness}, and also accounts for all independent $t_{ij}$ provided the manifold has no nontrivial topological structure. 

Inserting this form into \eqref{gaugedZ2def} we find
\be
\qquad Z' \sim \sum_{\{\sig_i,\lam_i\}} \exp\le(\beta\sum_{\vev{ij}} \lambda_i \sig_i \lam_j \sig_j\ri)
\ee
where the $\sim$ means that we are ignoring overall state-independent prefactors. Redefining each spin as $\sig_i \to \lam_i \sig_i$ we see that there is no dependence on $\lam_i$, and we obtain 
\be
Z' \sim \sum_{\{\sig_i\}} \exp\le(\beta\sum_{\vev{ij}} \sig_i \sig_j\ri) = Z
\ee
i.e. this is equivalent to the original ungauged model \eqref{ungauged}, as asserted above. 

\item In the second way of doing this calculation, we explicitly integrate out the original spins $\sig_i$ by noting that we can redefine $t_{ij} \to t_{ij} \sig_i \sig_j$ in \eqref{gaugedZ2def} to find that the $\sig_i$ dependence vanishes and we have only: 
\be
Z' = \sum_{\{\mathrm{flat}\;t_{ij}\}}\exp\le(+\beta \sum_{\vev{ij}} t_{ij}\ri) \ . 
\ee
We now need to sum over only flat $t_{ij}$. This can be done efficiently by writing a projector onto $F_p = 1$, i.e.
\be
\sum_{\{\mathrm{flat}\;t_{ij}\}} = \sum_{\{ t_{ij} \}} \prod_{p} \ha(1 + F_p) = 2^{-N}\sum_{\{ t_{ij}\}} \prod_{p} \sum_{\tilde{\sig}_p = -1}^{+1} F_p^{\ha(1 + \tilde{\sig}_p)}
\ee
where in the last two expressions the sum over $t_{ij}$ is now unconstrained, and in the last equality we have introduced a new variable valued on plaquettes $\tilde{\sig}_p = \pm 1$ to represent the projector onto zero flux in a useful manner. Ignoring overall constant prefactors in $Z'$ we thus find
\be
Z' = \sum_{\{t_{ij},\tilde{\sig}_p\}}\prod_{p}\le(\prod_{l \in p} t_{l}^{\ha(1 + \tilde{\sig}_p)}\ri) \prod_{l} e^{\beta t_l} \label{newZ} 
\ee
The partition function factorizes over the $t_l$ on independent links. It is thus possible to integrate out $t_l$ explicitly. The probability density for each $t_l$ depends on the two $\tilde{\sig}_p$ on the plaquettes on either side of it. Let us denote the spins on those two plaquettes as $\tilde{\sig}_p$ and $\tilde{\sig}_{p \perp l}$. We see that if $t_l = +1$ then the contribution to the sum is independent of $\tilde{\sig}_p \tilde{\sig}_{p \perp l}$ and is always $e^{\beta}$, whereas if $t_l = -1$ the contribution to the sum alternates in sign depending on the sign of $\tilde{\sig}_p \tilde{\sig}_{p \perp l}$ and has magnitude $e^{-\beta}$. Assembling the pieces we find:
\be
Z' = \sum_{\{\tilde{\sig}_p\}} \prod_{l} \le(e^{\beta} + e^{-\beta} \tilde{\sig}_p \tilde{\sig}_{p \perp l}\ri)
\ee
This is the partition sum of a dual Ising model with spins living on the plaquettes (or equivalently on the sites of the dual lattice). To compute the dual temperature $\tilde{\beta}$ we seek to write
\be
\le(e^{\beta} + e^{-\beta} \tilde{\sig}_p \tilde{\sig}_{q}\ri) = f(\beta) e^{\tilde{\beta} \tilde{\sig}_{p} \tilde{\sig}_{q}}
\ee
where $f(\beta)$ is a $\tilde{\sig}$-independent prefactor. Taking ratios and products of this expression for the two possible choices $\tilde{\sig}_{p} \tilde{\sig}_{q} = \pm 1$ we find
\be 
\tilde{\beta} = -\ha \log \tanh \beta \qquad f(\beta) = \sqrt{\tanh\beta}
\ee
Thus we find that
\be
Z' \sim \sum_{\{\tilde{\sig}_p\}}\exp\le(\sum_{\vev{pq}} \tilde{\beta} \tilde{\sig}_p \tilde{\sig}_q\ri) \qquad \tilde{\beta} = -\ha \log \tanh \beta
\ee
This is the usual statement of Kramers-Wannier duality on the 2d square lattice. Here $\vev{pq}$ now represents links on the dual lattice. 
\een 
\subsection{Kramers-Wannier duality with next-to-nearest neighbour interactions}

We now consider applying a similar technique to the 2d Ising model with next-to-nearest neighbour couplings as in \eqref{kappaham}:
\be
H[\sigma] = -\beta \sum_{\vev{ij}} \sig_i \sig_j - \kappa \sum_{[ij]} \sig_i \sig_j
\ee
We will now gauge the symmetry as we did above. The gauged version of the model requires us to pick a way to connect two sites that are diagonally coupled using the links on the original lattice and then populate the links with appropriate $t_{ij}$. There are multiple natural ways to do this for the diagonal lattice; they all result in the same partition function once we sum over all flat $t_{ij}$, but a choice of any one pattern will generically break some lattice symmetries at intermediate stages. To preserve all of the symmetries at all stages in the calculation we will sum over the possibilities in the action, i.e. for the coupling $\sig_{x,y}\sig_{x+1,y+1}$ we will gauge it in two distinct ways:
\be
\sig_{x,y}\sig_{x+1,y+1} \to \frac{1}{2}\le(\sig_{x,y} t_{(x,y),(x,y+1)} t_{(x,y+1),(x+1,y+1)} \sig_{x+1,y+1} + \sig_{x,y} t_{(x,y),(x+1,y)} t_{(x+1,y),(x+1,y+1)} \sig_{x+1,y+1} \ri)
\ee
For notational simplicity we will refer to this whole combination as $\sig_i tt_{[ij]}\sig_j$. Thus we find the gauged Hamiltonian to be
\be
H'[t,\sig] = -\beta \sum_{\vev{ij}} t_{ij}\sig_i \sig_j - \kappa \sum_{[ij]} tt_{[ij]} \sig_i \sig_j
\ee
We may now proceed to integrate out the original spins and introduce the dual as above. The analogue of \eqref{newZ} becomes
\be
Z' = \sum_{\{t_{ij},\tilde{\sig}_p\}}\prod_{p}\le(\prod_{l \in p} t_{l}^{\ha(1 + \tilde{\sig}_p)}\ri) \prod_{l} e^{\beta t_l + 2\ka\sum_{\delta} t_l t_{l+\delta}} \label{newZ2} 
\ee
where the sum over $\delta$ refers to the links $l+\delta$ that neighbour a given link at $l$. Note that $\kappa$ now introduces an interaction between the $t_{ij}$ variables and it is no longer possible to integrate them out explicitly and obtain an exact Hamiltonian for the dual spins $\tilde{\sig}_p$ in closed form.  

To proceed, we may attempt to explicitly integrate out the $t_l$ in a mean-field treatment. To that end we write
\be
t_l t_{m} = (t_l - \vev{t_l} + \vev{t_l})(t_m - \vev{t_m} + \vev{t_m}) 
\ee
and then expand out the right hand side, assuming (as usual in mean-field) that the fluctuations $t - \vev{t}$ are small and ignoring terms that are second order in $\sO(t-\vev{t})$. We then find
\be
t_l t_m \approx \vev{t_m}t_l + \vev{t_l} t_m - \vev{t_l}\vev{t_m}
\ee
Now the partition function \eqref{newZ2} can be written as
\be
Z' = \sum_{\{t_{ij},\tilde{\sig}_p\}}\prod_{l} t_{l}^{\ha(1 + \tilde{\sig}_p) + 
\ha\le(1 + \tilde{\sig}_{p \perp l}\ri)}  e^{\beta t_l + 2\ka\sum_{\delta} t_l t_{l+\delta}} \label{newZ2l} \ . 
\ee
To obtain a mean-field partition function $Z_{\mathrm{mf}}$ we replace $t_l t_{l+\delta}$ in \eqref{newZ2l} with its mean-field approximation to obtain
\be
Z_{\mathrm{mf}} = \sum_{\{t_{ij},\tilde{\sig}_p\}}\prod_{l} t_{l}^{\ha(1 + \tilde{\sig}_p) + 
\ha\le(1 + \tilde{\sig}_{p \perp l}\ri)}  e^{\beta t_l + 2\ka\sum_{\delta} (\vev{t_{l+\delta}}t_l + \vev{t_l} t_{l+\delta} - \vev{t_l}\vev{t_{l+\delta}})} \ . 
\ee
We can now integrate out the $t_l$ explicitly on each link, where the effect of the interaction is approximated by the terms of the form $t_l\vev{t_{l+\delta}}$. We find
\be
Z_{\mathrm{mf}} = \sum_{\{\tilde{\sig}_p\}}\prod_{l} (e^{B_l} + \tilde{\sig}_p \tilde{\sig}_{p \perp l} e^{-B_l})e^{- 2\ka\sum_{\delta} \vev{t_l}\vev{t_{l+\delta}}} \qquad B_l = \beta + \ka\sum_{\delta} \vev{t_{l+\delta}} \ . \label{Zmf} 
\ee
To close this set of equations, we compute $\vev{t_l}$ in the approximation above: 
\be
\vev{t_l} = \frac{e^{B_l} - \tilde{\sig}_p \tilde{\sig}_{p \perp  l} e^{-B_l}}{e^{B_l} + \tilde{\sig}_p \tilde{\sig}_{p \perp  l} e^{-B_l} } \label{tvev} 
\ee
Note that as $B_l$ depends on the neighbouring $\vev{t_{l+\delta}}$, this should be viewed as a kind of discrete ``partial difference equation'' for the variable $\vev{t_l}$, where this PDE depends on the values of the $\tilde{\sig}_p$. In principle one can solve this PDE for the $\vev{t_l}$ and then insert the result back into \eqref{Zmf} to obtain a dual Hamiltonian for the $\tilde{\sig}_p$. In practice this is not possible for finite $\ka$, so even the mean-field treatment cannot be explicitly performed.  It is clear that the effective Hamiltonian will contain non-local couplings. Further approximation schemes are possible -- e.g. one could expand order by order in $\ka$ on top of this mean field treatment -- but we do not pursue this further here.

\section{Experimental details} \label{app:exp} 
In this Appendix we provide some details on the experimental results from the machine learning approach of Section \ref{MLapproach}. 
\subsection{Neural network training} 

  A rough measure of the uncertainty of our results can be obtained from the spread of the learned $\tilde{\beta}$ points in Figure \ref{fig:beta-betaTilde}, which grows as we approach the phase transition at $\beta_c \approx 0.44$. Interestingly, the method does not perform reliably for $\beta > \beta_c$, when the original frame is in the symmetry-broken phase. This is somewhat reminiscent of known difficulties 
in learning parameters of Hamiltonians at high $\beta$ (see e.g. Appendix B of \cite{Haah:2021nzn}) and deserves further study.

Further, from Figure \ref{fig:beta-betaTilde} we observe that the initialization of $\beta$ has an impact on the frame that is chosen. We find that $\beta_0=0.4$ almost always drifted away from the original frame as indicated by the absence of green points on the $\beta = -\tilde\beta$ line. 
Note for some $\beta$ (e.g., $\beta=0.35$), we see some suboptimal runs resulting in the final $\beta$ far away from the dual or original frame. These have high loss and can be easily identified as failed runs. 

\begin{figure*}[htp!]
    \centering
    \includegraphics[width=1.0\linewidth]{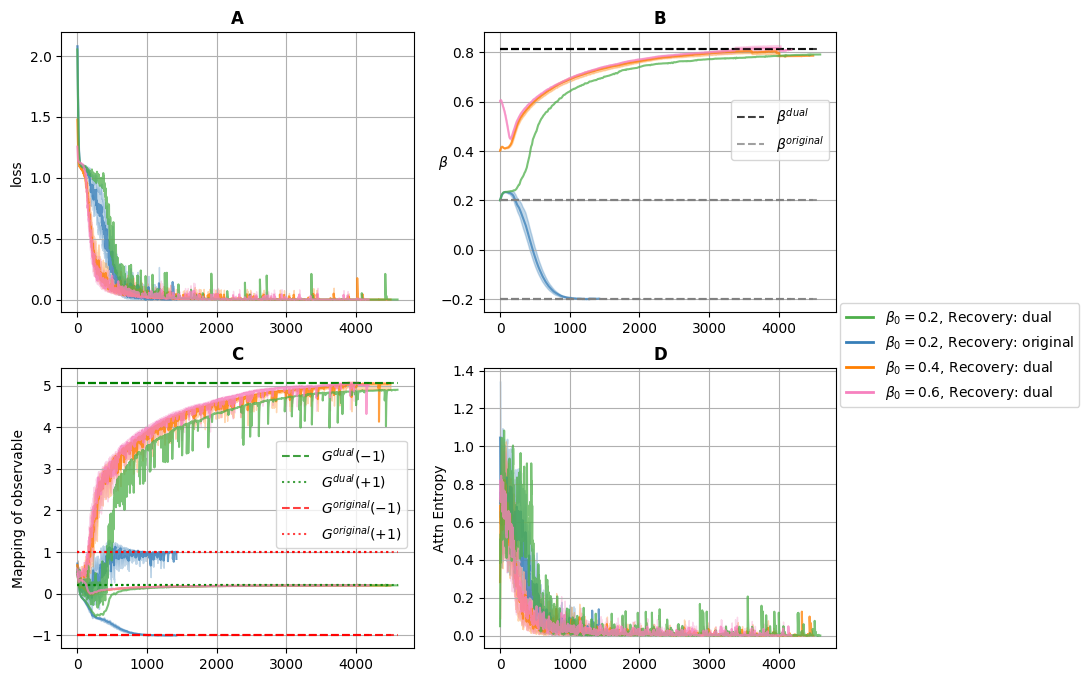}
    \caption{Training progress for runs from $\beta=0.2$, grouped by $\beta_0$ and $\tilde\beta^*$ to showcase the trajectory of various metrics. 
    We show exponentially smoothed moving average of the following metrics: (A) Loss, (B) $\beta$, (C) Mapping of observables, (D) Entropy of $\text{Gumbel-Softmax}(\theta_1)$ For (B) and (C) we denote theoretically expected values in original and dual frames by the dashed lines. }
    \label{fig:loss_trajectory}
\end{figure*}

Figure~\ref{fig:loss_trajectory} shows runs from $\beta=0.2$ grouped by $\beta_0$ and $\tilde\beta^*$,  illustrating how the training progresses under different scenarios. 
For the seeds where either the dual or original frame is recovered, the loss goes to 0. 
Further, we track the entropy of $\text{Gumbel-Softmax}(\theta_1)$ to assess how the algorithm is weighing each feature. 
A value of 0 corresponds to a strong preference for one out of the seven input links.
% We also show the training progress of the seed that yielded a non-optimal $\beta$ that ends at a higher loss than the rest. \NI{I do not think this last seed run is there, right?} 

\begin{figure*}[htp!]
    \centering
    \includegraphics[width=\linewidth]{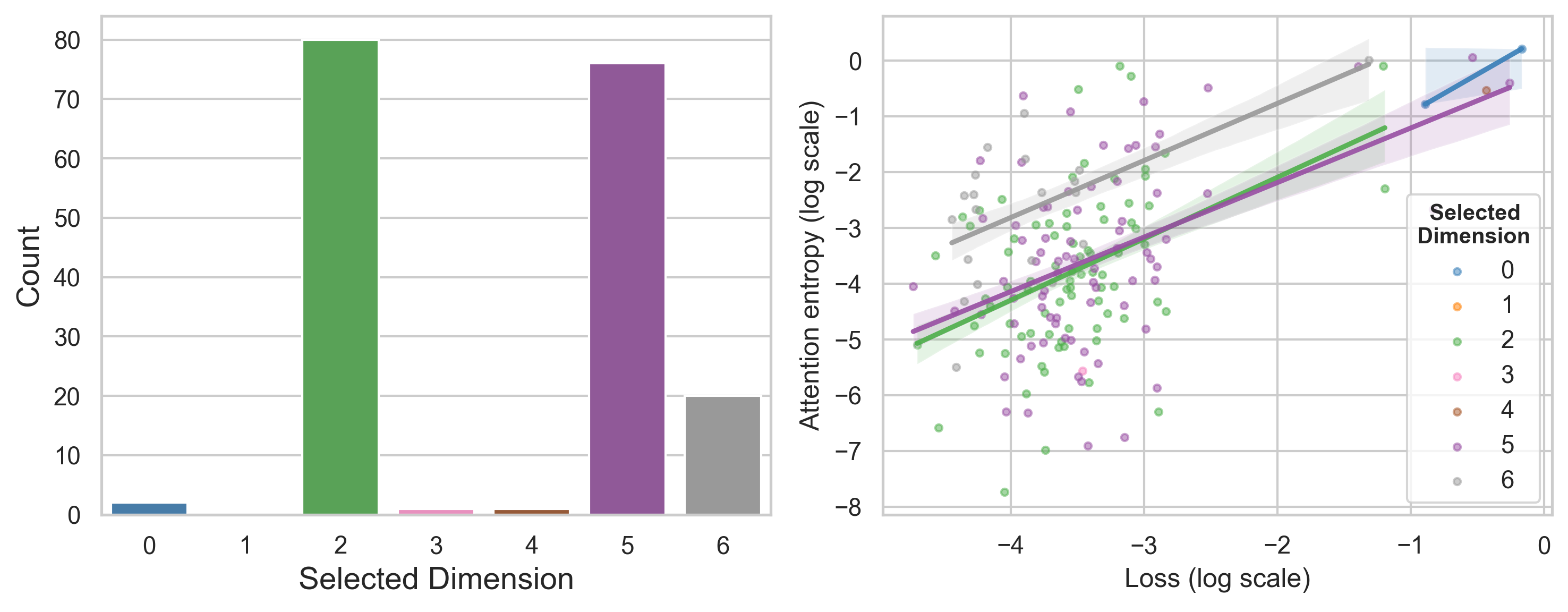}
    \caption{(left) Corresponding to Figure~\ref{fig:beta-betaTilde}, we plot the frequency with which the dimension of the feature vector $\mathbf{f}_{\vev{ij}}$ was selected. Note that $2$ and $5$ pertain to the dual frame, while $6$ relates to the discovery of the original frame. 
    (right) Loss values (log scale) and attention entropy (log scale) are positively correlated such that lower loss increasingly prefers a single dimension of feature vectors. Note that both the loss and attention entropy are very low on the three features corresponding to the dual and original frames, as expected for a successful run.
    }
    \label{fig:selected_dim}
\end{figure*}

Finally, in Figure \ref{fig:selected_dim} we see which link is selected as a map to the observable, using the numbering in Figure \ref{fig:linkMap}; links 2 and 5 correspond to a mapping to the dual lattice (and are found when we recover a map to the dual frame) and link 6 corresponds to recovering the original link (and are found when we recover the original frame). The small amounts of other dimensions correspond to failed runs which generically have a higher loss. 

\vspace{-15pt}

\subsection{Training hyperparameters}
\label{sec:training}

Our models are all implemented in PyTorch~\cite{paszke2017automatic}. 
We used the Adam~\cite{kingma2014adam} optimizer with the learning rate of $0.005$ for $\beta$ and $0.01$ for $\theta$. 
Moreover, we used the early stopping criterion to stop the training if the loss didn’t improve over 200 epochs. 
We ran the sampler in each experiment to generate 1000 samples for the lattice. 
We ran the training for a maximum of 25000 epochs, and our runs took about 1-3 hours each.
Our experiments are run on the lattice size of $8 \times 8$.

\subsection{Loss landscape in ordered phase}
\begin{figure*}[htp!]
    \centering
    \includegraphics[width=\linewidth]{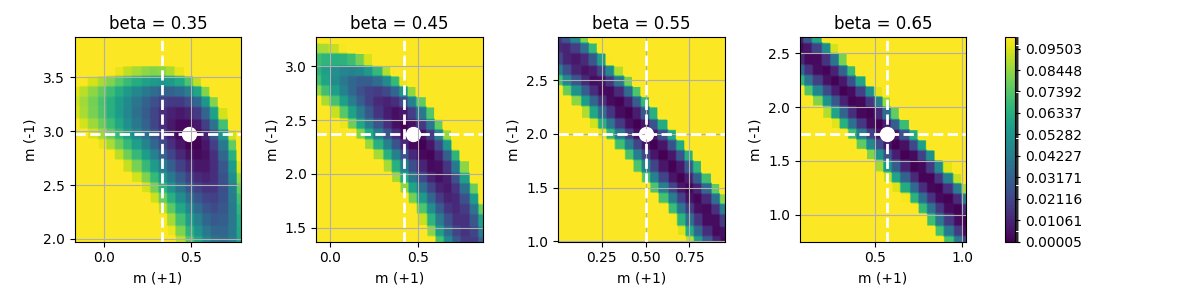}
    \caption{The formation of valleys beyond the phase transition in which the minimal losses lie creates a problem, resulting in a continual drift along the valleys once low loss values are reached. }
    \label{fig:valleys}
\end{figure*}

In the ordered phase (i.e. for $\beta > \beta_c \approx 0.44$), the proposed framework encounters challenges as the loss function develops a valley structure, obstructing convergence to a unique optimum as shown in Figure \ref{fig:valleys}. This formation of the loss valley is attributed to the specific choice of small-distance features picked, which seems to work well only within the disordered phase. 

\subsection{Scaling results}

\begin{figure}
    \centering
    \includegraphics[width=\linewidth]{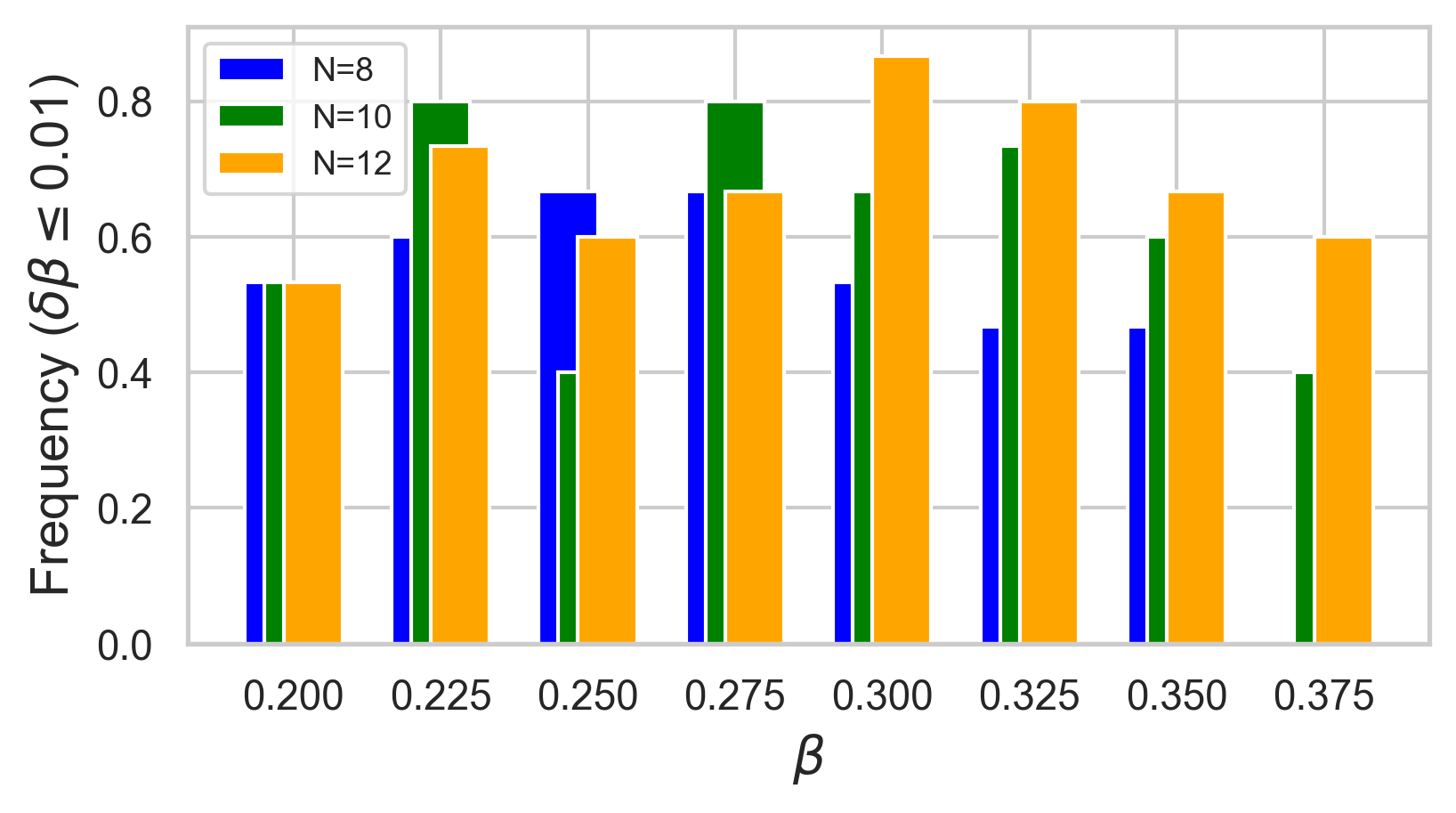}
    \caption{Fraction of times relevant beta was recovered within the tolerance of 0.01 as the lattice size is increased. }
    \label{fig:scaling}
\end{figure}

Figure~\ref{fig:scaling} shows the fraction of instances in which either $\tilde{\beta}$, $\beta$, or $-\beta$ were successfully recovered. We observe that as the lattice size increases to $10 \times 10$ and $12 \times 12$, the recovery rate improves.

\end{appendix}

\bibliographystyle{utphys}
\bibliography{all}
\end{document}